\documentclass[12pt]{iopart}
\eqnobysec

\usepackage{graphicx}
\usepackage{color}

\begin{document}

\title{Random matrix models for phase diagrams}

\author{B Vanderheyden$^1$ and A D Jackson$^2$}

\address{$^1$ Department of Electrical Engineering and Computer 
Science (B28) and SUPRATECS, University of Li\`ege, Belgium }

\address{$^2$ The Niels Bohr International Academy, The Niels Bohr
 Institute, Blegdamsvej 17, DK-2100 Copenhagen \O, Denmark}

\ead{B.Vanderheyden@ulg.ac.be}

\begin{abstract}

We describe a random matrix approach that can provide generic and
readily soluble mean-field descriptions of the phase diagram for a
variety of systems ranging from QCD to high-$T_c$ materials.  Instead
of working from specific models, phase diagrams are constructed by
averaging over the ensemble of theories that possesses the relevant
symmetries of the problem.  Although approximate in nature, this
approach has a number of advantages.  First, it can be useful in
distinguishing generic features from model-dependent details.  Second,
it can help in understanding the `minimal' number of symmetry
constraints required to reproduce specific phase structures.  Third,
the robustness of predictions can be checked with respect to
variations in the detailed description of the interactions.  Finally,
near critical points, random matrix models bear strong similarities to
Ginsburg-Landau theories with the advantage of additional constraints
inherited from the symmetries of the underlying interaction. These
constraints can be helpful in ruling out certain topologies in the
phase diagram. In this Key Issue, we illustrate the basic structure of
random matrix models, discuss their strengths and weaknesses, and
consider the kinds of system to which they can be applied.

\end{abstract}

\pacs{12.38 Aw, 74.25.Dw, 71.10.Fd}

\section{Introduction}

Determining the phase diagram of a system is a central issue in many
areas of physics. Knowledge of the thermodynamic state of a system as
a function of externally controlled variables (e.g., temperature,
density, or magnetic field) often provides important macroscopic
information and can serve to help in identifying correlations among
its microscopic constituents.  A further determination of the nature
of these correlations, which can be a consequence either of local
interactions or of the cooperative behavior of a large number of
constituents, is central to understanding the physical laws that
govern the system.

In this Key Issue, we discuss the thermodynamics of two seemingly
different types of systems: strongly interacting matter at extreme
temperature and density, and high-$T_c$ compounds (including cuprates
and ferropnictides) at finite temperature and doping. Strongly
interacting matter is governed by Quantum Chromodynamics (QCD). For
QCD with two flavors of massless quarks, a chiral and/or deconfining
phase transition is expected at a temperature of $T \sim 160~$MeV for
zero quark chemical potential~\cite{HEMCGC93} or at lower temperatures
for a chemical potential as large as $\mu \sim
300~$MeV~\cite{Halasz1998}.  In this latter regime, attractive quark
interactions may lead to a 'color-superconducting'
phase~\cite{Barrois1977, Frautschi1978, Bailin1984, Alford1998,
  Rapp1998, Rajagopal2000}.  The high density end of the phase diagram
is relevant for the physics of dense stars~\cite{Rajagopal2000, Itoh1970,
  Alford2008}; the lower densities are relevant for relativistic
heavy-ion collisions~\cite{Rajagopal1999} where it is hoped that
unambiguous signatures of this phase transition can be observed.

The energy scales appropriate for the description of high-$T_c$
materials are evidently much (i.e., roughly $10^{10}$ times) smaller.
For low doping, these materials exhibit an antiferromagnetic phase
(AF) with transition temperatures of $\sim 150~$K to $400~$K.  For
higher doping, these materials exist in a superconducting phase (SC)
with a transition temperature of tens of degrees kelvin. This
schematic picture can be refined to obtain a very rich phase structure
with coexisting magnetic and pairing correlations and a variety of
complex phenomena including stripes and pseudogaps~\cite{Dagotto94,
  Scalapino06, Lee2006, Carlson2008, Ni2008, Chu2009}. The basic
mechanism for pairing is not well understood, and identifying the
manner in which AF and SC phases dominate or coexist in the
competition for the same electrons may be helpful in eliminating
certain scenarios~\cite{Fernandes2010, Fernandes2010a, Vavilov2010,
  Vorontsov2010}.

Both QCD and high-$T_c$ materials are characterized by strong
correlations among their microscopic constituents, and there is no
indication in either case of a suitable ``small parameter'' that
could lead to a meaningful perturbative description.  Both systems
would appear to be too complicated to permit analytic descriptions.
Even numerical approaches are either extremely challenging or
impossible as a consequence of the fundamental fermion sign
problem~(e.g., See \cite{Dagotto94, Loh1990, Barbour1998, Hands2007}).
For each of these systems, numerical investigations yield highly
oscillatory integrals that can prevent their simulation using
techniques based on importance sampling.  Given this situation, it may
be useful to begin from the observation that these phase transitions
are associated with the breaking of a symmetry (e.g., the staggered
magnetization of AF breaks the spin rotational symmetry of the
system).

In this Key Issue, we describe how an approach based on random matrix
theory (RMT) can provide useful information regarding the possible
phase diagrams for a variety of systems. Central to this approach is
the identification of relevant symmetries and their consequences for
the macroscopic properties of the system.

The basic ideas of random matrix theory (RMT) were introduced by
Wigner when studying the complex spectra of intermediate and heavy
nuclei~\cite{Wigner51,Wigner55,Wigner67}.  His rationale can be
summarized as follows: Although the detailed properties of the lowest
energy states of nuclei could be understood by means of model
Hamiltonians, there was no hope that this approach could explain the
excited resonances observed with neutrons of energy up to several
hundred eV.  These states were far too numerous and too closely spaced
to permit detailed modelling.  It was deemed to be both possible and
more meaningful to determine the {\it statistical} properties of an
ensemble of states, such as their density in a given energy interval
or the distribution of their energy spacings and widths.  This was
achieved by constructing Hamiltonians with a structure that was
directly dictated by the quantum physics of nuclei, but which
otherwise contained matrix elements that were drawn at random on an
appropriate distribution. The statistical properties of the highly
excited resonances were then obtained by performing suitable ensemble
averages analytically.

The central elements of this construction were the random and the
deterministic parts of the Hamiltonians.  Here, the random part of the
Hamiltonian was not intended to describe randomly fluctuating
processes or statistical disorder.  Randomness was rather used as a
statistical instrument: One considered a large ensemble of nuclei,
each having a different Hamiltonian, and determined the properties of
a group of nuclei by computing ensemble averages. The deterministic
part of the Hamiltonian provided constraints on the resulting
statistical properties.  Initially, this deterministic structure was
that dictated by the time-reversal symmetry of the strong interaction,
which implies that the Hamiltonian can be written as a real symmetric
matrix.  By choosing statistical distributions so that ensemble
averages were representative of the vast majority of the nuclei
considered, the resulting properties were shown to be independent of
the detailed dynamics of the nuclear interactions. Only the underlying
symmetries were important.

These early ideas have evolved considerably since, and random matrix
theory is now applied in many branches of nuclear, condensed matter,
and particle physics.  Still, in each case, symmetry plays a central
role. This Key Issue will focus on the application of random matrix
theory to the study of phase diagrams in systems where different
symmetries compete thermodynamically.  As in the early work of Wigner,
random matrix theory is well suited for this task when the considered
systems exhibit many intermixing energy states whose detailed
properties are largely unknown or very difficult to determine.
Starting from model Hamiltonians constrained by symmetries and generic
mechanisms, the results are expected to be broadly representative of
an ensemble of systems respecting these constraints.  Ensemble
averaging will eliminate details specific to individual Hamiltonians,
and the features of the resulting phase diagram should thus be
sufficiently robust that they will emerge in ``almost all'' specific
cases.

Before considering this specific application, we wish to clarify the
context in which it has been developed.  Nowadays, random matrix
theory encompasses a vast domain of research which has been reviewed
in many works. (A non-exhaustive list is available in~\cite{review1,
  review2, review3}.)  The consequences of symmetry constraints can be
studied at two different levels --- microscopic and macroscopic.
Microscopic investigations deal with the correlations of the
eigenenergies of the system on the scale of the average
spacing. Examples of RMT predictions at this level include the
spectral properties of nuclear resonances, atomic energy levels, sound
waves in quartz crystals, transmission modes in quantum dots, and the
eigenvalues of the QCD Dirac operator. One of the most important
microscopic results is universality.  When RMT applies, eigenvalue
correlations are not only dictated by the symmetries of the system but
are also independent of the details of the statistical distributions
used to generate the ensemble averages.  (See, e.g., section 7
of~\cite{review2}).

The second, macroscopic level of description deals with the
consequences of the underlying symmetries on the global state of the
random matrix system. This is the level of the models considered
below.  Here, the central quantity is the partition function, from
which the phase competition can be determined as a function of the
variables of the theory such as temperature or chemical potential.
The resulting phase diagram is a direct consequence of the symmetries
of the system, and it is of interest to study their robustness with
respect to variations of the coupling constants of the theory.  At
this stage, it is useful to note one detail of the random matrix
approach: Matrix elements are drawn independently in such a way that
all basis states are treated equally.  Such `democratic treatment'
naturally implies that random matrix models are mean-field models.

These two levels of description are closely related to one another.
For instance, in QCD with two massless flavors, it is observed that
the smallest eigenvalues of the Dirac operator in vacuum accumulate
near the origin of the spectrum.  This accumulation is directly
related to the spontaneous breaking of the chiral symmetry, as
expressed by the Banks-Casher relationship which provides a direct
connection between the average spectral density near the origin and
the chiral order parameter~\cite{Banks1980}.  Another strong
relationship can be seen in the fact that many macroscopic properties
are `inherited' from symmetry constraints imposed at the microscopic
level. As will be shown below, the random matrix partition function
near the critical lines can be expanded as a power series in the order
parameters.  In this sense, RMT bears a strong resemblance to
Ginsburg-Landau theory.  In contrast to the Ginsburg-Landau approach,
however, the coefficients in the random matrix expansion {\it
  cannot\/} be freely chosen but must satisfy relationships resulting
from the form of the microscopic random matrix interactions. These
constraints can preclude the occurrence of some symmetry breaking
patterns.  Thus, RMT can be used to determine the minimum number of
symmetries which must be imposed at the microscopic level in order to
reproduce a given phase structure.

In this Key Issue, we describe the basic steps required to construct
random matrix models for phase diagrams.  In the interests of
readability, we have not attempted to provide technical details.
Emphasis is rather placed on the ongoing simplification that arises in
this construction, from the many variance parameters and complicated
block structure needed to describe the random matrix interactions at
the microscopic level to the few parameter ratios and elementary
functional forms that characterize the thermodynamic potential.

We start with two basic models in section~\ref{s:two-models}.  The
first describes chiral symmetry breaking in QCD, which is related to a
quark-antiquark order parameter of the form $\langle
\psi^\dagger\psi\rangle$ (in Euclidean conventions).  The second is
the case of an attractive phonon-like interaction, which produces
pairing states with an order parameter of the form
$\langle\psi\psi\rangle$. In section~\ref{s:QCDvsTc}, we summarize our
previous work on QCD and high-$T_c$ superconductivity~\cite{VanJac00a,
  VanJac00b, VanJac01, VanJac03, VanJac05, VanJac09}. Although different in
nature, these systems exhibit phase structures with striking
similarities in their symmetry breaking patterns. In particular, we
discuss the conditions required for the occurrence of either a
tetracritical or a bicritical point, corresponding respectively to the
presence or the absence of a phase with mixed broken symmetries. In
section~\ref{s:oxy}, this discussion is extended to a tentative model
for iron arsenide compounds with an $s_\pm$ order parameter, and we
discuss the possibility that this system can have a state with
coexisting magnetic and superconducting
correlations. Section~\ref{s:conclusions} presents final remarks
regarding the random matrix approach to phase diagrams as well as a
selected list of open problems.

\section{Two basic models}
\label{s:two-models}

\subsection{Chiral symmetry breaking in QCD}
\label{ss:chiral}

As a first illustration of the random matrix approach, we consider a
model for chiral symmetry breaking in QCD with two light flavors,
leading to an order parameter of the form $\langle
\psi^\dagger\psi\rangle$. Working in Euclidean space and assuming a
zero vacuum angle, the QCD partition function for fixed quark chemical
potential, $\mu$, and temperature, $T$, is given as
\begin{eqnarray}
 Z(\mu,T) = e^{-\Omega(\mu,T)/T} = \int {\cal D}\psi^\dagger{\cal D}\psi
{\cal D}A \: e^{-S_E},
\end{eqnarray}
where $S_E$ is the Euclidean action
\begin{eqnarray}
S_E  = \int_0^{1/T} dx_0 \int d^3 x \left({1\over 2 g^2}\: \mathrm{Tr}
F_{\mu\nu}^2 - \sum_{f=1}^{N_f} \psi^\dagger_f \left(  {\cal
 D}_{\mathrm{QCD}} + m_f + \mu \gamma_0 \right)\psi_f\right).
\label{QCDaction}
\end{eqnarray}
Here, $\psi_f$ and $\psi_f^\dagger$ describe quark and antiquark
fields of mass $m_f$ and $N_f=2$ is the number of flavors. Quarks
interact via the exchange of gluon fields,
\begin{eqnarray}
A_\mu=\sum_{a=1}^3 T_a A_{\mu a},
\end{eqnarray}
where $T_a$ are the generators of the $SU(N_c)$ color algebra (with $N_c=3$
throughout this section), and the coupling arises via the Euclidean
Dirac operator 
 \begin{eqnarray}
    {\cal D}_{\mathrm{QCD}} = \gamma_\mu(\partial_\mu + i A_\mu),
 \end{eqnarray}
where the $\gamma_\mu$ are hermitean. ($\gamma_0$ is the matrix coupled
to the time derivative $\partial_0$.)  The dynamics of the gluon
fields is contained in the quadratic term, $\mathrm{Tr}
F_{\mu\nu}^2/(2 g^2)$, where
\begin{eqnarray}
F_{\mu\nu} =  \partial_\mu A_\nu - \partial_\nu A_\mu + i \left[A_\mu,A_\nu
\right],
\end{eqnarray}
the trace is carried over colors, and $g$ is a coupling constant. As
described in the introduction, the resulting interactions are too
complex to be solved exactly, and one must resort to approximations.

The random matrix approach starts from symmetry considerations: In the
chiral limit, $m_f \to 0$, the QCD partition function is invariant
under SU($N_f$)$\times$SU($N_f$).  Nevertheless, chiral symmetry is
spontaneously broken to SU$_V$($N_f$) in the vacuum~\cite{Hoo86}.  In
the early 90's, it was discovered that chiral symmetry breaking could
be seen in the dynamics of the small eigenvalues of the Dirac operator
under fluctuations of the gauge field configurations.  It was already
known from the work of Banks and Casher~\cite{Banks1980} that the
chiral order parameter, $\langle \psi^\dagger\psi\rangle$, can be
expressed as an accumulation of the eigenvalues near zero
virtuality. (The actual relationship will be made explicit below in
equation~(\ref{BK}).) In 1992, Leutwyler and Smilga derived further
sum rules for the spectrum of the Dirac operator, which are dominated
by the eigenvalues near zero virtuality~\cite{LeuSmi92}. In 1993,
Shuryak and Verbaarschot~\cite{ShuVer93} introduced a random matrix
model for chiral symmetry breaking, where the QCD partition function
in vacuum is approximated as
\begin{eqnarray}
\fl
Z_{\mathrm{RMT}}(\mu=0,T=0)  =  \int\,{\cal{D}} W \,\,\prod_{f = 1}^{N_f}\,\,
{{\cal D}}\psi_f^*\,{{\cal D}} \psi_f^{\phantom{*}} \, \exp\left[ \sum_{f= 1}^{N_f} \,
 \psi^*_f \,\left( {\cal D}_\mathrm{RMT} + m_f \right)\psi_f
 \right]\nonumber\\ \times\exp\Big(-  n 
 \Sigma^2\,{\rm Tr}[W W^\dagger]\Big).
\label{ZRMTXsB}
\end{eqnarray}
In this approach, it is expected that the spectrum of the Dirac
operator, ${\cal D}_\mathrm{RMT}$, has the same properties as ${\cal
  D}_\mathrm{QCD}$ provided the model satisfies the same global
symmetries as QCD.  In particular, in the chiral limit $m_f=0$ the
Dirac operator should anticommute with $\gamma_5$. Working in the
basis of the eigenstates of $\gamma_5$, ${\cal D}$ must therefore have
the block structure
\begin{eqnarray}
{\cal D}_{\mathrm{RMT}}= i \left(
\begin{array}{cc}
0 & W\\
W^\dagger & 0 \\
\end{array}
\right),
\label{Dirac}
\end{eqnarray}
where $W$ is an $n\times n$ complex matrix which models the QCD
interactions. The choice of complex elements corresponds to the chiral
unitary ensemble (chGUE), which is relevant for fermions in the
fundamental representation.  There exists two other ensembles, the 
chiral orthogonal (chGOE) and chiral symplectic (chGSE) ensembles, 
which are discussed in~\cite{Verprl94}.

The dynamics of the interactions are now materially simplified.  The
quadratic term $\sim F_{\mu\nu}^2$ in the action (\ref{QCDaction}) is replaced
by drawing the matrix elements of $W$ on a Gaussian distribution with
zero mean and inverse variance $\Sigma$. Hence, an ensemble average is
made over gauge field configurations. The corresponding number of
degrees of freedom is proportional to $n$, which serves as the volume
of the system and is to be taken to infinity at the end of the
calculations (i.e., the thermodynamic limit). Note that the matrix
elements $W_{ij}$ are drawn on the same distribution for all $i$ and
$j$.  In other words, the random matrix interactions are independent
of the choice of basis states.  As a result, no spatial range can be
specified.  The only scale introduced is that of the interaction
strength, $\sim 1/\Sigma$.

Given the form of equation (\ref{Dirac}), the spectrum of ${\cal
 D}_{\mathrm{RMT}}$ has the desired properties.  Because the Dirac
operator anticommutes with $\gamma_5$, eigenvalues occur in pairs
$\pm i \lambda$ with eigenstates $u_i$ and $\gamma_5\,u_i$, so that the
spectrum is symmetric about $\lambda=0$. Chiral symmetry is however
spontaneously broken in the ground state. The chiral order parameter
can be obtained from the Banks-Casher relationship~\cite{Banks1980} as
\begin{eqnarray}
|\langle\psi^\dagger\psi\rangle| = \lim_{\lambda\to 0}\:\lim_{m_f\to
  0}\:\lim_{n\to\infty}\frac{\pi \rho(\lambda)}{2n},
\label{BK}
\end{eqnarray}
where
\begin{eqnarray}
 \rho(\lambda) = 
\left\langle \sum_i \delta(\lambda - \lambda_i)
\right\rangle
\end{eqnarray}
is the average spectral density.  Note that the different limits in
(\ref{BK}) do not commute. They must be taken in the order indicated
since eigenvalues accumulate near $\lambda = 0$ as $n$ is taken to
infinity.  (In particular, it can be seen that reversing the
$\lambda\to 0$ and $n\to \infty$ limits would give zero, since
$\rho (0) = 0$ for any finite $n$ as a consequence of the $\pm
\lambda$ symmetry noted above.)  

Further properties of the spectrum were shown to match those of QCD
(for eigenenergies $\lambda \ll \Lambda_\mathrm{QCD}$).  First, the
model of equations (\ref{ZRMTXsB}) and (\ref{Dirac})  was shown to obey the QCD
Leutwyler-Smilga sum rules~\cite{LeuSmi92,VerZah93}. Second, direct
comparisons with data from numerical simulations of QCD on a lattice
showed that the spectrum of the Dirac operator simulated at the scale
of the average energy spacing (the `microscopic level density') agrees
with the random matrix predictions within computational
uncertainties~\cite{lattice1,lattice2, lattice3, lattice4,
  lattice5, lattice6}.  All together, these results demonstrated that
the eigenvalue correlations near $\lambda = 0$ are a direct
consequence of the underlying chiral symmetry and do not depend on the
specifics of the interactions.  In that sense, the correlations are
universal.  Of course, QCD could have been an ``exceptional'' theory
with spectral correlators different from the ensemble average.  The
fact that it is not the case provides encouragement that random matrix
arguments can also offer insight into its  macroscopic properties.

Chiral symmetry breaking can be studied at a macroscopic level
with the aid of the thermodynamic potential. The chiral random matrix
model can be solved exactly by standard methods. There are three
steps.  First, the matrix elements $W_{ij}$ are integrated over to
give~\cite{JacVer96}
\begin{eqnarray}
Z_{\mathrm{RMT}}(0,0) = \prod_{i = 1}^{N_f}\,\, {{\cal
    D}}\psi_i^\dagger\,{{\cal D}} \psi_i^{\phantom{*}} \, e^Y \, ,
\end{eqnarray}
where $Y$ is an effective quartic interaction potential given as
\begin{eqnarray}
Y = - {1\over n \Sigma^2}\,\sum_{f,g,i,k}
 \psi^{f\dagger}_{Lk} \psi^f_{Ri} \psi^{g\dagger}_{Ri} \psi^g_{Lk} +
\sum_{f,i}
 m_f (\psi^{f\dagger}_{Li} \psi_{Li}^f + \psi^{f\dagger}_{Ri} \psi_{Ri}^f),
\label{Y-XsB}
\end{eqnarray}
Second, each fermion bilinear is expressed as combinations of
squares which can then be linearized by means of a Hubbard-Stratonovitch
transformation
\begin{eqnarray}
 e^{A Q^2} \sim \int dx \exp\left(- \frac{x^2}{4 A} - Qx\right),
\end{eqnarray}
which introduces an auxiliary variable, $x$.  Focusing on the
term relevant for chiral condensation, only one such auxiliary
variable, $\sigma$, is considered, and the partition function becomes
\begin{eqnarray}
 Z_{\mathrm{RMT}}(0,0) \sim \int d\sigma\, e^{- 2 n N_f \Omega(\sigma)}, 
\label{ZafoOmega}
\end{eqnarray}
with a thermodynamic potential 
\begin{eqnarray}
 \Omega(\sigma) = {\Sigma^2\,\sigma^2\over 2} - \log(|\sigma + m|),
\label{Omega}
\end{eqnarray}
where we have assumed $m_1=m_2=m$ for simplicity.  Interestingly, the
function $\Omega(\sigma)$ is even in the limit $m=0$, reflecting
chiral symmetry.  For the final step, the integral on the right side of
equation (\ref{ZafoOmega}) can be solved exactly in the thermodynamic
limit $n\to\infty$ by the saddle-point method, as $\lim_{n\to\infty}
(1/2 n N_f)\log Z_{\mathrm{RMT}} = - \min_{\sigma} \Omega(\sigma)$. In the
chiral limit $m\to 0$, the minimum of $\Omega(\sigma)$ is simply given as
\begin{eqnarray}
 \sigma_0 = 1/\Sigma.
\end{eqnarray}
The corresponding chiral condensate in vacuum is then
\begin{eqnarray}
 \langle\psi^{\dagger}\psi\rangle & = & \lim_{m\to 0}\lim_{n\to\infty}
            {1\over 2 n N_f }\,\frac{\partial \log Z}{\partial m} =
             \Sigma.
\end{eqnarray}
Note again that the thermodynamic limit must be taken first to
produce a meaningful result. In conclusion, the vacuum state has 
a non-zero order parameter as a result of the spontaneous breaking of
chiral symmetry.

This analysis can be extended to finite temperature and density in
order to describe the complete phase diagram. Technically, $T$ and the
chemical potential, $\mu$, are introduced in the Dirac operator by the
fermion Matsubara frequencies $\omega_n = \mu + (2 n +1) i \pi T$
(where $n$ is an integer).  In other words, ${\cal D}$ is replaced by
${\cal D} + \gamma_0 \omega_n$, and the partition function is summed
over the $\omega_n$'s~\cite{JacVer96,Halasz1998}.  The thermodynamic
potential can again be determined by following the procedure described
above.  The result is simply
\begin{eqnarray}
\Omega(\sigma) = A \sigma^2 - \sum_{n,\pm}
\log\left((\pm(\sigma + m) -\mu)^2+(2 n+1)^2\pi^2T^2\right),
\label{OmegaRMCsB}
\end{eqnarray}
where $A$ is a constant proportional to $\Sigma^2$. Note the structure
of the logarithm term, which can be rewritten as
$\sum_{n,\pm}\log(E_\pm^2+(2 n+1)^2\pi^2T^2)$, where $E_\pm = \pm (\sigma +
m)-\mu$ represent the quark ($+$) and antiquark ($-$) single
quasiparticle energies in a fixed auxiliary field, $\sigma$.

A simplified potential can be obtained by restricting the frequency sum
to the two smallest frequencies, $\mu + i \pi T$ and $\mu - i \pi
T$~\cite{JacVer96}.  The resulting potential can then be minimized
analytically and gives a polynomial gap equation for
$\sigma_0(\mu,T)$,
\begin{eqnarray}
{A \sigma_0} = \sum_\pm \frac{\sigma_0 + m \pm \mu}{(\sigma_0 + m \pm
  \mu)^2 + \pi^2 T^2},
 \label{OmegaSimple} 
\end{eqnarray}
which can be solved exactly in order to describe the phase diagram in the
$(\mu,T)$ plane.  

 \begin{figure}
   \begin{center}
     \begin{tabular}{cc}
      \includegraphics[width=0.48\textwidth]{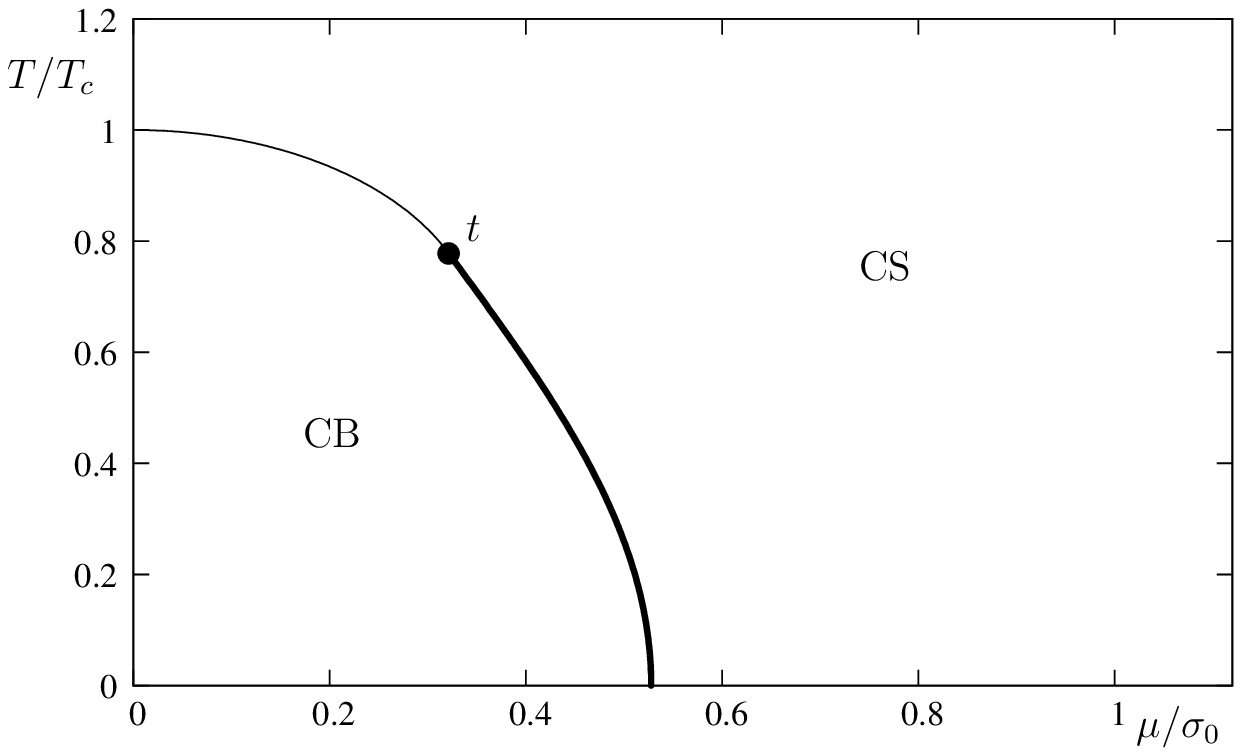}
&
      \includegraphics[width=0.48\textwidth]{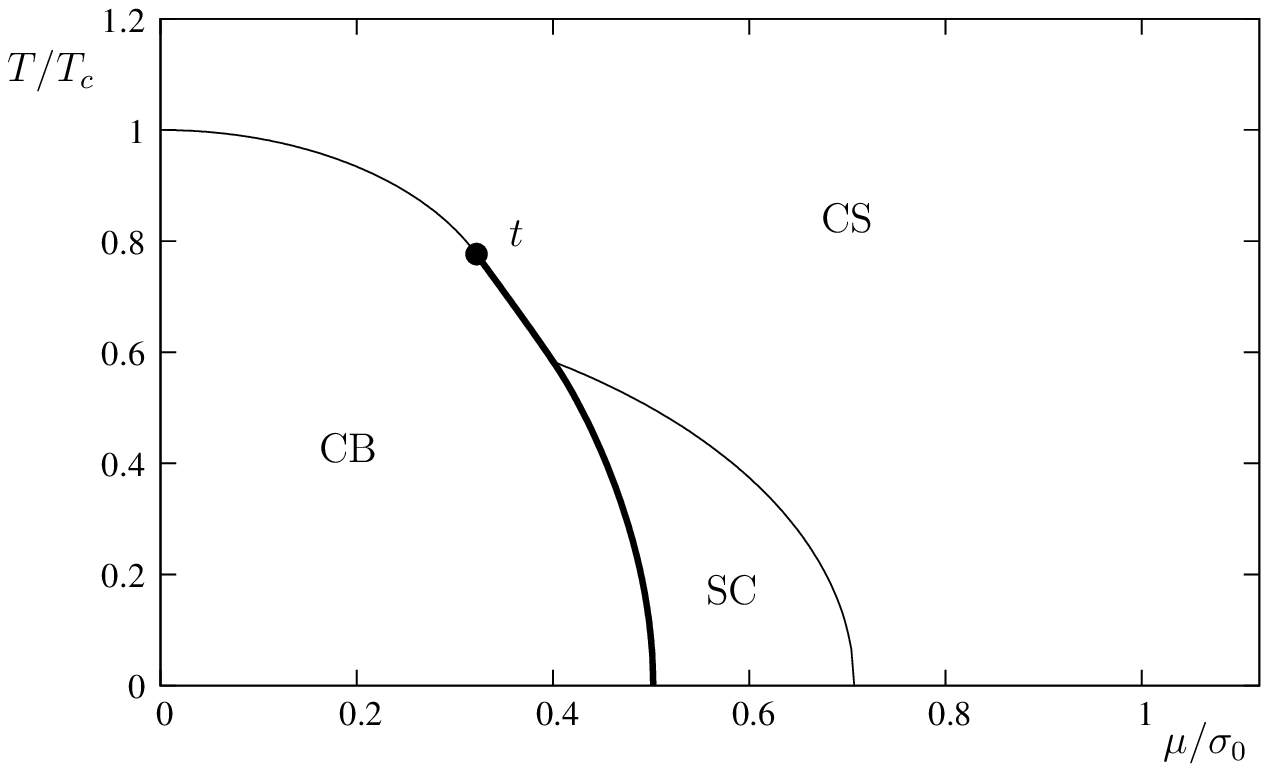}
\\
      (a) & (b) \\
      \\
      \includegraphics[width=0.48\textwidth]{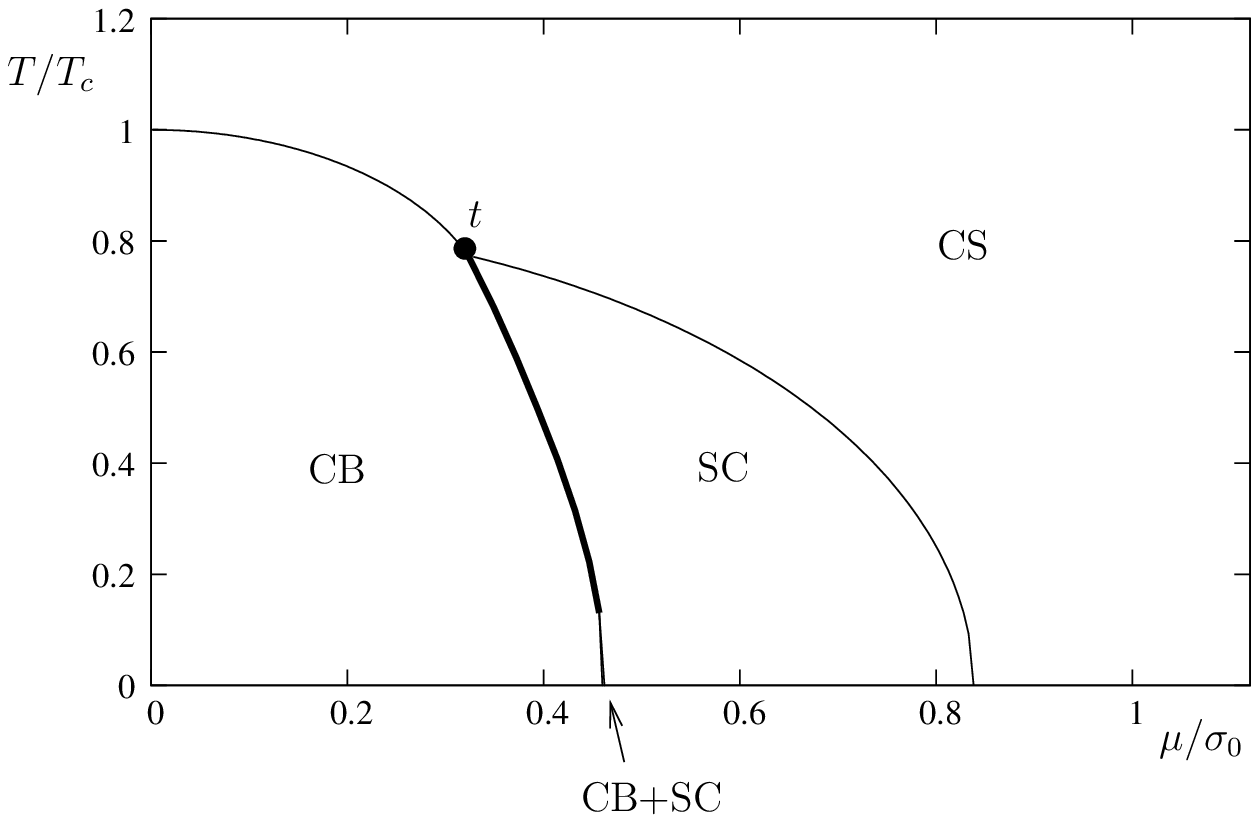}
&
      \includegraphics[width=0.48\textwidth]{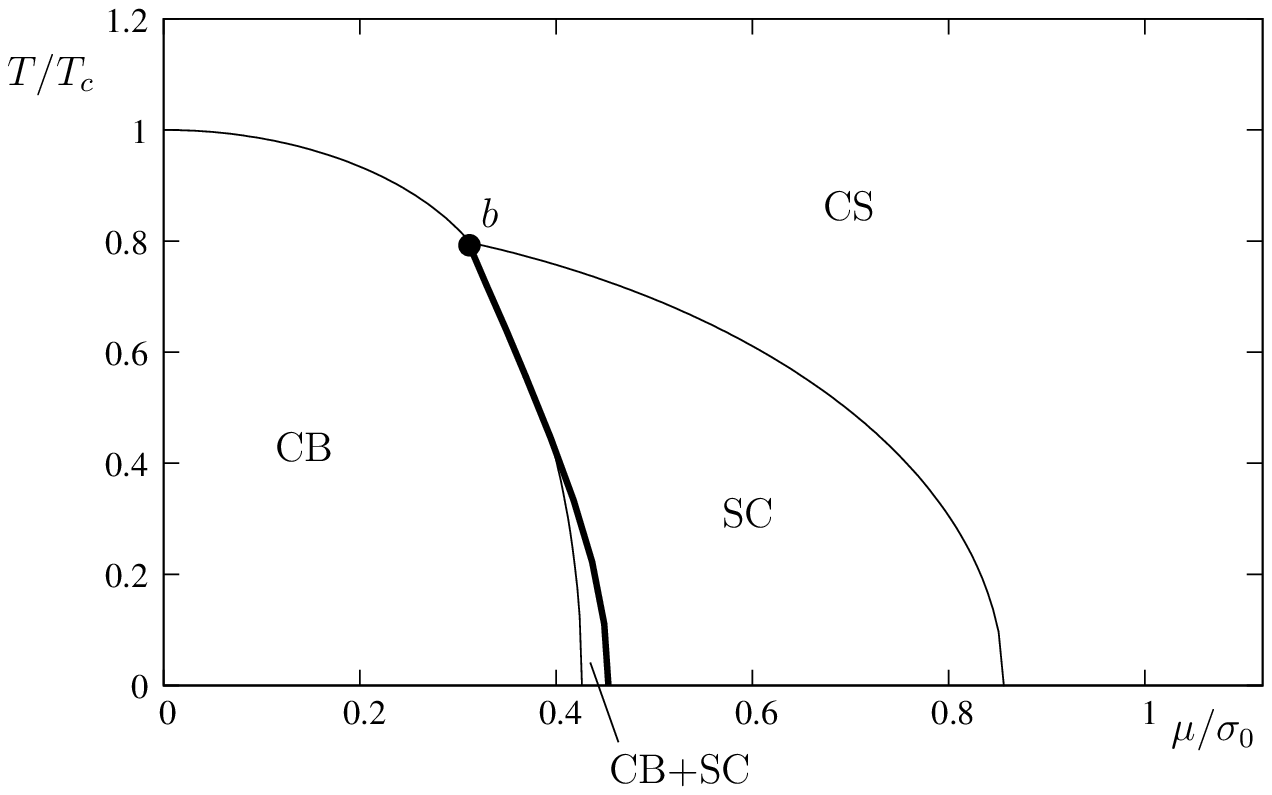}
\\
      (c) & (d) \\
      \\
      \includegraphics[width=0.48\textwidth]{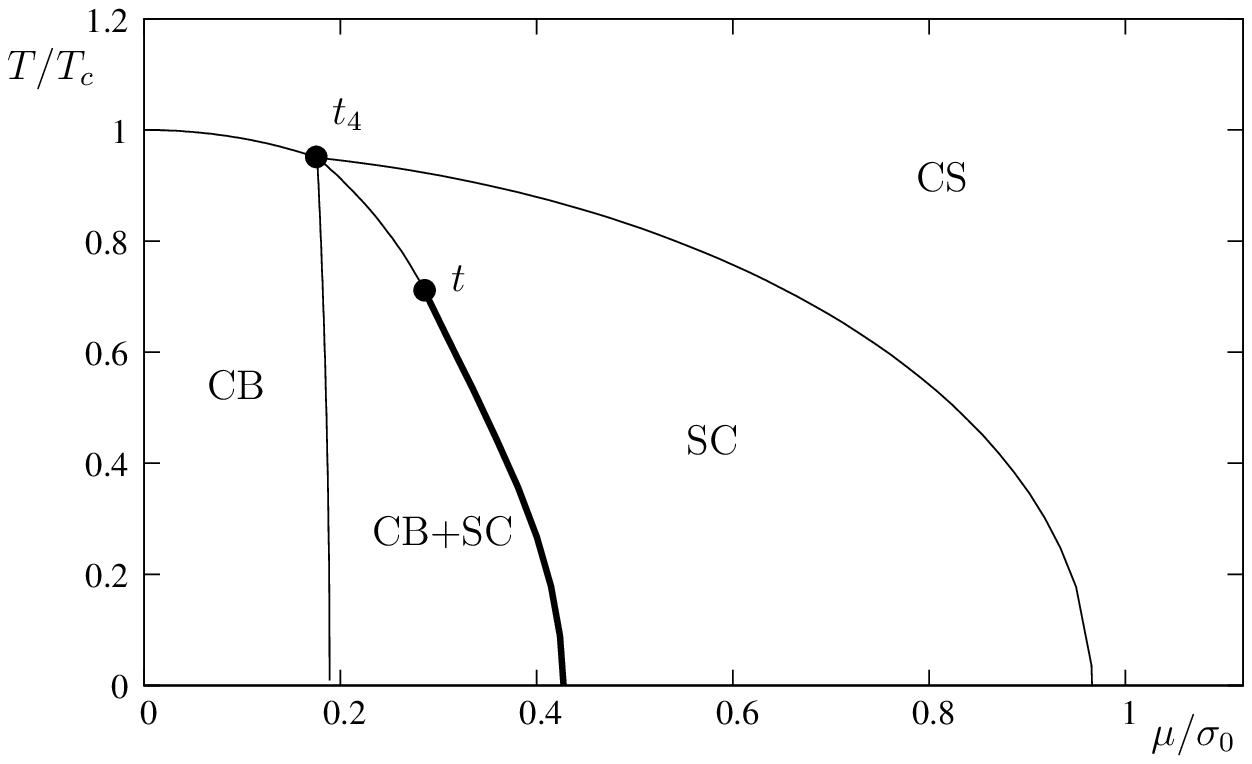}
&
      \includegraphics[width=0.48\textwidth]{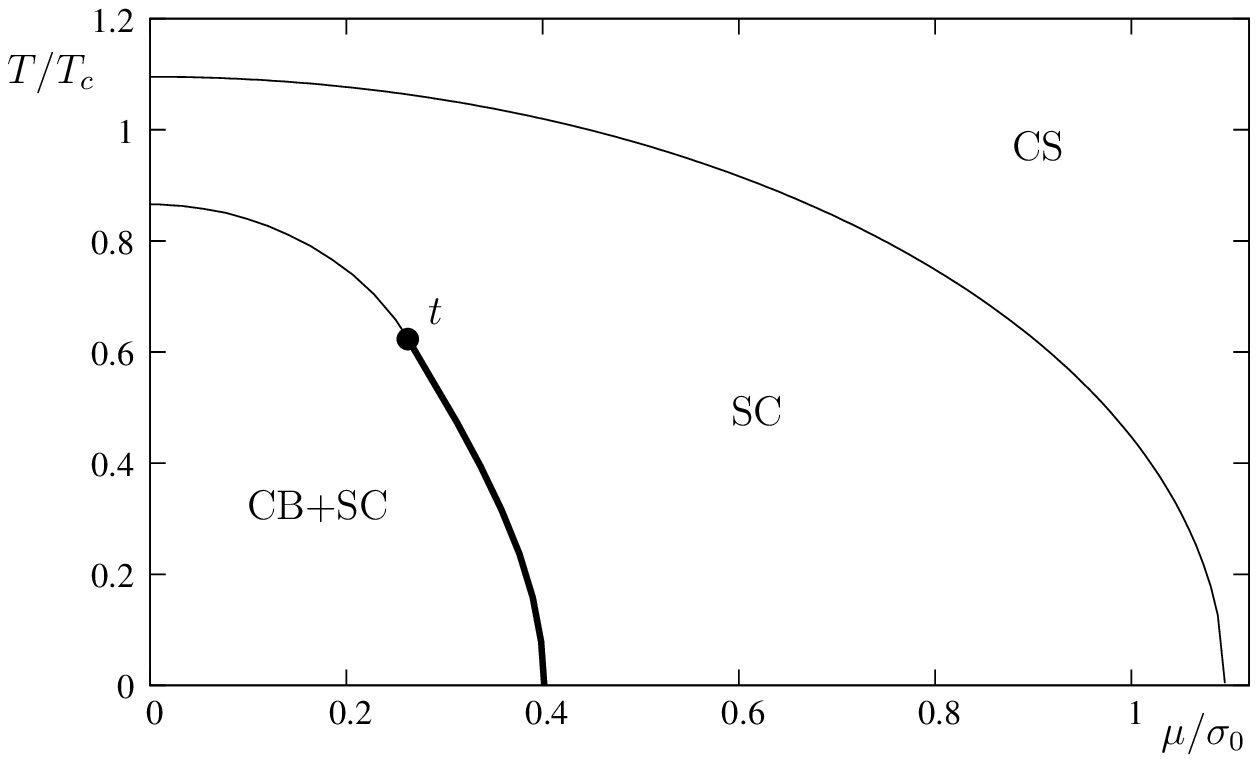}
\\
      (e) & (f) \\
     \end{tabular}
   \caption{Figure (a) shows the phase diagram for
     the chiral symmetry breaking model of section~\ref{s:two-models}.
     This phase structure also shows one of the six topologies that
     can be realized in the random matrix model for the competition
     between chiral symmetry breaking and quark pairing, see
     section~\ref{s:QCDvsTc}. CB is the chiral broken phase, CS the
   chiral symmetric one, SC the superconducting phase, and CB+SC the
   mixed broken symmetry phase. (This is a phase with an homogeneous
   coexistence of the two orders.) First- and second-order transitions
   are depicted by thick and thin lines, respectively. $t$ denotes a
   tricritical point, $t_4$ a tetracritical point, and $b$ a
   bicritical one. For the diagrams (a)-(e), $T_c$ is the transition
   temperature at $\mu=0$ and $\sigma_0$ is the value of the chiral
   field in vacuum. For diagram (f), $T_c$ and $\sigma_0$ assume the
   same expressions (i.e., $T_c=\sigma_0=\sqrt{3/B}$) even though they
   no longer correspond to a transition temperature or a chiral
   condensate. From top left to bottom right, the panels correspond to
   $\alpha=0.3$, $\alpha=0.75$, $\alpha=1.054$, $\alpha=1.1$,
   $\alpha=1.4$, and $\alpha=1.8$. Note that the phase structure (c)
   contains a small wedge of mixed broken symmetry which is separated
   from CB by a second-order line and from SC by a first-order
   line. The figures are adapted
   from~\cite{VanJac00b}. \label{f:QCD-pd}}
   \end{center}

 \end{figure}

The resulting phase diagram is shown in figure~\ref{f:QCD-pd} (a) in
the chiral limit $m \to 0$. (The other phase structures will be
discussed in section~\ref{s:QCDvsTc}.) In vacuum, $\sigma_0 \neq 0$ and chiral
symmetry is spontaneously broken. For small values of $\mu$, chiral
symmetry is restored at a finite temperature via a continuous
(second-order) phase transition. By contrast, chiral symmetry
restoration along the zero temperature axis is realized by means of a
first-order phase transition, with discontinuities in both the
pressure and the density of the system. In this respect QCD bears a
strong analogy to an antiferromagnet, where an increase of the applied
magnetic field induces an abrupt vanishing of the staggered
magnetization of the ground state.  In QCD, the chemical potential is
an intensive quantity whose increase leads to a variation of the
baryon density $\langle \psi^\dagger \gamma_0 \psi\rangle$.  However,
increasing $\mu$ also causes reduction of the chiral order parameter,
$\langle \psi^\dagger\psi \rangle$.  Similar to the abrupt vanishing
of the staggered magnetization under an applied magnetic field,
$\langle \psi^\dagger\psi \rangle \to 0$ through a first-order
transition.  The rest of the phase diagram must be continuously
related to the transitions observed at $T=0$ and $\mu=0$.  As shown in
figure~\ref{f:QCD-pd} (a), the second-order line starting from the
$\mu=0$ axis continuously joins the first-order line starting from the
$T=0$ axis. The two lines meet at a tricritical point, $t$, located at
finite values of $T$ and $\mu$.

The macroscopic results are clearly of a different nature from those
obtained at the microscopic level. It is important to understand the
role played by the global symmetries at the macroscopic level. How
good are the random matrix predictions for the phase diagram?  Do they
have a universal character? To answer these questions, two regions
must be distinguished --- near the critical lines and away from them.

In the critical regions, the random matrix potential of equation
(\ref{OmegaRMCsB}) resembles that typically obtained in a
Ginsburg-Landau approach. Along the second-order critical line,
$\Omega(\mu,T)$ can be expanded as a series of powers of $\sigma$ to
give
\begin{eqnarray}
\Omega(\mu,T) = \Omega_0(\mu,T) + a(\mu,T)\, \sigma^4 +
b(\mu,T)\,\sigma^6 + \ldots,
\end{eqnarray}
where $a(\mu,T)$ and $b(\mu,T)$ are {\it known} functions of $\mu$ and
$T$.  In this form, random matrix theory resembles a mean-field
$\phi^4$ theory.  Furthermore, the coefficient $a(\mu,T)$ vanishes at
the tricritical point, and the theory exhibits the critical exponents
of a mean-field $\phi^6$ theory.  This result would also have been
obtained by a more traditional Ginsburg-Landau approach.  It is
important, however, to recognize that the construction of the
random matrix model at the deeper microscopic level of the
interactions fixes the coefficients $a(\mu,T)$ and $b(\mu,T)$
uniquely.  This is not the case in Ginsburg-Landau theory.  This fact
has consequences on the thermodynamic competition between different
phases and will be discussed further in section~\ref{s:QCDvsTc}.

Away from the critical regions, the shape of the phase diagram must be
interpreted with care.  The phase structure was established on the
basis of a severely truncated sum over Matsubara frequencies.  This is
a rough approximation, which was shown to be equivalent to a mere
rescaling of the temperature through a monotonic
function~\cite{VanJac01}.  (Note that this truncation is not necessary
if the phase diagram is to be considered numerically.)  Such a
transformation shifts the phase boundaries and the critical points in
the $(\mu,T)$ plane, but it preserves the overall shape of the phase
diagram. Hence, the topology of the phase diagram --- the relative
positions of the different phases --- is a direct consequence of the
underlying symmetry and is a robust prediction. The predicted order of
a given phase transition is more delicate to assess. As stressed
earlier, the random matrix approach ignores most of the dynamics of
the interactions. A more detailed description can affect the energy
balance between the different states of the system and lead to local
modifications of the order of a phase transition, e.g. change a
second-order line into a weak first-order one.  This level of
approximation is however perfectly acceptable as such corrections are
to be expected in any mean-field approach. Indeed, the order parameter
is weak in the critical regions, and it is thus fairly sensitive to
the inclusion of additional thermal or quantum fluctuations.

\subsection{Phonon mediated superconductivity}
\label{ss:ph}

As a second example, we consider a random matrix model for a system of
electrons with a longitudinal phonon interaction leading to a pairing
condensate $\sim \langle \psi\psi \rangle$. This model will be
compared to results expected from mean-field approaches based on the
BCS theory in order to illustrate the method and to highlight
differences with respect to the more complete microscopic approach.
The random matrix model was motivated by a more extensive study
including antiferromagnetism and superconductivity~\cite{VanJac09},
the results of which will be discussed in section~\ref{s:QCDvsTc}.
Once again, some calculational details have been omitted in order to
focus on the basic structure of the model.

We wish to reproduce the basic structure of the partition function
\begin{eqnarray}
 Z_{\mathrm{ph}}(\mu,T) = \int {\cal D}\phi\, {\cal D}\psi^\dagger {\cal D}\psi
\, e^{-\int_0^{1/T} dx_0
   (H_{0-el}- \mu N+H_{0-ph} + H_\mathrm{int}) }, 
\label{Zph}
\end{eqnarray}
where $\phi(x)$ is a real field describing longitudinal phonons,
$\psi$ describe electrons, $H_{0-el} - \mu N$ is their free energy,
$H_{0-ph}$ is the Hamiltonian of free phonons, and the electron-phonon
coupling is given as~(see, e.g.,~\cite{phonon-field-theory-book})
\begin{eqnarray}
H_\mathrm{int} = g \sum_\alpha\int d^3 x\,
\psi^\dagger_\alpha(x)\,\psi_\alpha(x)\phi(x),
\label{Hint-ph}
\end{eqnarray}
with $\alpha$ the spin index and $g$ a real coupling constant. It is
implicitly assumed that the phonon interaction decreases sufficiently
rapidly with distance to justify the contact form of
$H_\mathrm{int}$. The phonon interaction is attractive and competes
with Coulomb repulsion; for electrons near the Fermi surface, the
phonon interaction dominates and tends to produce $s-$wave electrons
pairs with an order parameter
\begin{eqnarray}
 m_{SC-s} = \left\langle \sum_{\mathbf{p},
   \omega_n,\alpha,\beta}\psi_\alpha(\mathbf{p},\omega_n)
 (i\sigma_2)_{\alpha,\beta}
 \psi_\beta(-\mathbf{p},-\omega_n)\right\rangle,
\label{pairingOPph}
\end{eqnarray}
where $i \sigma_2$ is the antisymmetric spin matrix. Pairing leads to
a reorganisation of the ground state of the system, and in particular,
leads to the opening of a gap in the quasiparticle spectrum. The gap,
$\Delta$, obeys the equation 
\begin{eqnarray}
 \Delta = \frac{|\lambda| T}{(2\pi)^3} \sum_n \int d^3 p \:\frac{
   \Delta}{((2 n+1) \pi T)^2+(\varepsilon(p) - \mu)^2+\Delta^2},
\label{gapeqph}
\end{eqnarray}
where $|\lambda|\sim g^2$, $n$ are integers, and $\varepsilon(p)-\mu \approx
v (p-p_F)$ is the energy of a single non-interacting electron near the 
Fermi surface. In this construction, it is implicitly assumed that the  
relevant states are centered around the Fermi
energy, $\mu$, in a narrow shell with an energy width $\sim 2 \omega_D$
where $\omega_D$ is the Debye energy for longitudinal phonons.

A random matrix formulation would naturally start with a partition
function of the form
\begin{eqnarray}
 Z_{\mathrm{RM-ph}}(\mu, T) = \int {\cal D}\psi^\dagger {\cal D}\psi
  {\cal D}H_{\mathrm{int}} P[H_{\mathrm{int}}] \, e^{-\psi^\dagger H \psi},
\label{ZRMph}\end{eqnarray}
where $H = H_0 - \mu N + H_\mathrm{int}$ is the sum of the
single-electron Hamiltonian and an hermitean term $H_\mathrm{int}$
describing the exchange of phonons. This term is chosen diagonal in
spin space. The term $P[H_{\mathrm{int}}]$ describes the dynamics of
the phonon fields. In order to construct an order parameter of the form
of equation (\ref{pairingOPph}), the following parametrization is
introduced: the full set of states is divided in two complementary
subsets $\{1,\alpha, i\}$ and $\{2,\alpha, i\}$ where the index 
$i$ denotes $(\mathbf{p},\omega_n)$ in the first 
subset and $(-\mathbf{p}, -\omega_n)$ in the second.  With
this parametrisation, the random matrix version of the order parameter
assumes the form
\begin{eqnarray}
 m_{SC-s} = \left\langle \sum_{i,\alpha,\beta} \psi_{1,i,\alpha}
 (i\sigma_2)_{\alpha, \beta} \psi_{2,i,\beta} \right\rangle.
\label{mSCsRM}
\end{eqnarray}
The interactions $H_\mathrm{int}$ are constrained by time-reversal
symmetry. In coordinate representation, the phonon field is real since
it carries no electric charge. In momentum representation, we thus
require that the Fourier components of the matrix elements satisfy the
complex conjugation constraint
\begin{eqnarray}
 ({\mathbf{p},\omega_n}|H_\mathrm{int}|\mathbf{q},\omega_m)=
(-{\mathbf{p},-\omega_n}|H_\mathrm{int}|-\mathbf{q},-\omega_m)^*.
\end{eqnarray}
Using the parametrization just introduced, it can be shown that this
is equivalent to imposing the following block
structure~\cite{VanJac09}~:
\begin{eqnarray}
 H_{\mathrm{int}} = \mathbf{1}_\mathrm{spin} \left(
\begin{array}{cc}
B & C\\
C^* & B^*\\
\end{array}
\right),
\label{HintRMph-block}
\end{eqnarray}
where the hermiticity of $H_\mathrm{int}$ requires that $C$ is complex
symmetric and $B$ is Hermitean. Because the form of the time-reversal
operator depends on the basis of states used, the block structure of
the random matrix $H_{\mathrm{int}}$ is associated with a specific
choice of basis of states (i.e., states in the momentum representation
related by time-reversal symmetry).

As in our first example from QCD, we wish to explore
the consequences of the block symmetry in
equation~(\ref{HintRMph-block}) while considerably simplifying the
dynamics of the phonon interactions.  First, we assume that the 
important states are those for which phonon exchange
dominates Coulomb repulsion, as in the microscopic theory.  As a second
approximation, we note that the detailed theory of
equation~(\ref{Zph}) contains a free phonon term $H_{0-ph}$ which is a
bilinear function of the field $\phi$. As in QCD, we replace this
term by a random matrix with elements drawn on a Gaussian
distribution with zero mean and fixed inverse variances, $\Sigma^2_B$
and $\Sigma_C^2$. This gives
\begin{eqnarray}
 P[H_{\mathrm{int}}] = \int {\cal D}B{\cal D}C 
\exp\left(-N (\Sigma_B^2 \mathrm{Tr} B B^\dagger + 
\Sigma_C^2 \mathrm{Tr} C C^\dagger)\right),
\end{eqnarray}
where $N$ represents the total number of states (over the two
subsets) and will be taken to infinity at the end of the
calculations. This form introduces the possibility of having different
strengths associated with scatterings from a region to itself
($\Sigma_B$) and between different regions ($\Sigma_C$).  A particular
choice can be made that reflects the known nature of the phonon
interaction for any given material, but this is not essential for what
follows. Except for this difference, no other scale is introduced.
With the parametrization introduced above, the resulting interaction
has an infinite range in momentum space and is thus a contact interaction 
in coordinate representation, as in equation (\ref{Hint-ph}).

To proceed, we again restrict the sum over Matsubara frequencies to
the two lowest terms, $\pm \pi T$. The free energy term
becomes
\begin{eqnarray}
\fl
\psi^\dagger (H_0 - \mu N) \psi=  \sum_{\pm,i,\alpha} \left( (\pm i \pi T - \mu) 
 \psi^\dagger_{1,\pm,i,\alpha} \psi_{1,\pm,i,\alpha} + 
(\mp i \pi T - \mu) \psi^\dagger_{2,\pm,i,\alpha} \psi_{2,\pm,i,\alpha} \right),
\label{HintRMph}
\end{eqnarray}
where, consistent with the parametrisation introduced above, states in
$\{1\}$ and $\{2\}$ have opposite Matsubara frequencies. (Here, the
Matsubara index $\pm$ is written out explicitly.)  The random
matrix model is solved following the three-step procedure described
above: First, the interaction matrix elements are integrated over,
yielding an effective potential $Y$ with quartic terms of the form
$\sim \psi^\dagger_i \psi_j \psi^\dagger_j \psi_i$ and $\sim
\psi^\dagger_i \psi_j \psi^\dagger_i \psi_j$.  (Here, only the random
matrix indices are shown.)  The second set of terms, which describe an
attractive interaction favoring the formation of pairing condensates
$\langle \psi_i\psi_i\rangle$, are of particular interest.  Second,
the quartic terms are linearized by means of Hubbard-Stratonovich
transformations. Third, a thermodynamic potential is calculated
exactly in the thermodynamic limit $N\to\infty$.  The minimum of this
potential describes the thermodynamic state of the system as a
function of $T$ and $\mu$. The entire procedure is described in detail
in~\cite{VanJac09}. Concentrating on the pairing channels, we find
\begin{eqnarray}
Z_{\mathrm{ph-RM}} \sim \int d\Delta d\Delta^* \exp(- 2 N \Omega)
\end{eqnarray}
with
\begin{eqnarray}
 \Omega(\Delta) = A |\Delta|^2 - {1\over 2} \log(\mu^2 + |\Delta|^2 +
 \pi^2 T^2),
\end{eqnarray}
where the auxiliary field $\Delta$ is complex and is associated with
the order parameter of equation~(\ref{mSCsRM}) while the constant $A$
is given as
\begin{eqnarray}
 A = {1\over \Sigma^{-2}_B + \Sigma^{-2}_C}.
\end{eqnarray}
Note the structure of $\Omega(\Delta)$, which is similar to that found
for QCD in equation~(\ref{OmegaRMCsB}). It contains a quadratic term $A
|\Delta|^2$ which represents the energy cost for creating a
condensate and a logarithmic term which can be rewritten as 
$\sum_\pm\log(E_\pm^2 + \pi^2 T^2)$ where $E_\pm = \pm
(\mu^2+|\Delta|^2)^{1/2}$.  This expression is reminiscent of the
quasiparticle energies in the microscopic theory, $\pm ((\varepsilon_p
-\mu)^2+|\Delta|^2)^{1/2}$ with $\varepsilon_p$ taken to zero.  It is 
hardly surprising to see this limit emerge in the random matrix framework 
since no electron energy was introduced in the single-electron Hamiltonian
$H_0$.

It is also interesting to consider the gap equation, which is given 
in the random matrix model as
\begin{eqnarray}
 \frac{\partial\Omega}{\partial\Delta^*} =  0 \quad\Longrightarrow\quad
A \Delta  =  {\Delta \over 2 (\mu^2 + |\Delta|^2 +
 \pi^2 T^2)}.
\label{gapeq-phRM}
\end{eqnarray}
Comparison with the gap equation (\ref{gapeqph}) of the microscopic
theory shows evident similarities, but (\ref{gapeq-phRM}) is polynomial and 
can thus be solved exactly: 
A paired state with $\Delta \neq 0$ is found
for $\mu^2 + \pi^2 T^2 \leq 1/(2 A)$, and the system is unpaired
($\Delta = 0$) otherwise.  As a result, at a
given density (i.e., at some fixed $\mu$), the system undergoes a continuous
transition from a paired to an unpaired state at $T=T_c =
\pi^{-1}(1/(2 A) - \mu^2)^{1/2}$. For small gaps, $\Omega$ can be
expanded as $\Omega(\Delta) \approx \Omega(0) + a(\mu,T) |\Delta|^2 +
b(\mu,T) |\Delta|^4$, where $a(\mu,T)$ and $b(\mu,T)$ are known
functions. Near $T_c$, we have $a \sim (T-T_c)$, so that the theory
gives the critical exponents of a mean-field $\phi^4$ theory.

The reader may be surprised that the random matrix gap equation
(\ref{gapeq-phRM}) does not share a well-known feature with the BCS
gap equation: In the zero temperature limit and for $\Delta \to
0$, the right side of (\ref{gapeqph}) contains a logarithmic
singularity for $\varepsilon(p)$ near the Fermi surface. As a result,
a solution with $\Delta \neq 0$ exists no matter how weak the strength
$\lambda$. In contrast, the random matrix gap equation admits a paired
solution only if $\mu^2 + \pi^2 T^2 \leq 1/(2 A)$. It follows that
weak interactions, corresponding to large inverse variances
$\Sigma_B^2$ and $\Sigma_C^2$ as well as a large value of $A$, favor a
paired state in only a restricted region of the $(\mu,T)$ plane and a
vanishing gap in the rest of the phase diagram. 

Comparing (\ref{gapeqph}) and (\ref{gapeq-phRM}), the 
absence of a logarithmic singularity in the  random matrix approach 
can be attributed to two factors: First, the truncation of the Matsubara 
sum, which prevents the construction of a Fermi-Dirac distribution, and 
second, the lack of an explicit reference to momentum in the right side 
of equation  (\ref{gapeq-phRM}), which prevents the construction of a density
 of states over a range of energies. As a result, for fixed $\Delta$,
 the random matrix model contains only two energy states, $E_\pm$,
 where a microscopic model would have energies $\varepsilon(p)$
 continuously distributed near the Fermi surface $\mu$. The gap
 equation is then polynomial and does not contain the logarithmic
 singularity.  The discrepancy with the BCS approach should not be
 considered as a weakness.  The vanishing of the random matrix gap
 occurs in regions of parameter space where the condensate 
 would otherwise be weak and fragile against the inclusion of
additional fluctuations (either quantum or thermal).  Determining the
fate of the gap in these cases clearly requires going beyond
mean-field, even in a BCS approach.  The level of approximation
offered by the random matrix approach thus appears reasonable.

\section{Phase diagrams in QCD and in high-Tc materials}
\label{s:QCDvsTc}

The nature of the correlations that arise when different symmetries
compete for the same degrees of freedom and the macroscopic
manifestations of this competition are matters of primary concern in
condensed matter systems.  The result is often a rich phase structure
that can depend sensitively on the coupling parameters of the theory.
We will now show how random matrix theory can be used to address these
questions by considering two apparently different systems --- QCD at
finite temperature and density and high-$T_c$ superconductors and
$d-$wave pairing as found in the cuprates.

For QCD, we consider the thermodynamics competition between chiral
symmetry breaking, described in \ref{ss:chiral}, and quark pairing,
which is analogous to the electron pairing discussed in \ref{ss:ph}.
Quark pairing might occur in dense and cold quark matter as a result
of the attraction provided by the strong interaction in the color
antitriplet
channel~\cite{Barrois1977,Frautschi1978,Bailin1984,Alford1998,
  Rapp1998}. The formation of quark pairs might lead to energy gaps as
large as $\Delta \sim 100~\mathrm{MeV}$ for quark chemical potentials
of the order of $300~\mathrm{MeV}$ and have direct consequences on the
phase structure of nuclear matter at high density. In the following,
we consider the case of QCD with two light flavors, for which the
dominant pairing order parameter is given as
\begin{eqnarray}
 \left\langle \psi_{f\alpha\sigma}(p)
 \psi_{g\beta\sigma^{'}}(-p)\right\rangle = \phi(p^2)\,
 \varepsilon_{fg}\,\varepsilon_{\alpha\beta3}\,\varepsilon_{\sigma\sigma^{'}},
\end{eqnarray}
where $p$ is a four-momentum, $f$ and $g$ are flavors, $\alpha$ and
$\beta$ denote colors, while $\sigma$ and $\sigma^{'}$ are spins. The
tensors $\varepsilon$ impose antisymmetric combinations of the
various indices. This order parameter breaks color symmetry from
$SU(3)$ to $SU(2)$ while preserving the  $SU_L(2)\times SU_R(2)$ flavor symmetry.

For high-$T_c$ materials, we consider a two-dimensional system of
electrons on a square lattice. There is a vast literature reporting a
rich and complex phase structure arising from the competition between
magnetic and pairing
correlations~\cite{Dagotto94,Scalapino06,Lee2006}.  Proposed order
parameters include $d-$wave pairing (see~\cite{Bulut1991,Monthoux1991}
and references therein), stripes~\cite{Carlson2008}, or correlations
arising from the pseudogap phenomenon~\cite{Loram93,Emery02}.  Here, we
investigate the competition between two of these possibilities.  The
first one is antiferromagnetism with an order parameter of the form
\begin{eqnarray}
 \mathbf{m}_{AF} = \left\langle \sum_{\mathbf{p}\,\omega_{n}\alpha\beta} 
\psi^{\dagger}_\alpha(\mathbf{p+Q},\omega_n)\,
\mbox{\boldmath$\sigma$}_{\alpha\beta}^{\phantom{\dagger}}\, 
\psi^{\phantom{\dagger}}_\beta(\mathbf{p},\omega_n)\right\rangle,
\label{AF-OP-microscopic}
\end{eqnarray}
where $\mathbf{p}$ are momenta in the first Brillouin zone,
$\mathbf{Q}= (\pi \hbar/a,\pi \hbar/a)$ is the AF ordering vector ($a$
is the lattice spacing), $\omega_n = (2 n +1 ) \pi T$ are fermion
Matsubara frequencies, $\alpha$ and $\beta$ are spin indices,
{\boldmath $\sigma$} are the spin Pauli matrices, and $\langle \ldots
\rangle = \mathrm{Tr}(\ldots e^{-\beta H})$ denotes a thermal
average. The second order parameter is associated with $d$-wave
pairing and is given as
\begin{eqnarray} 
m_{SC-d} = \left\langle \sum_{\mathbf{p},\omega_n} g(\mathbf{p})\,
\psi_\uparrow
(\mathbf{p},\omega_n)\,\psi_\downarrow(-\mathbf{p},-\omega_n)\right\rangle,
\label{SC-d-OP-microscopic}
\end{eqnarray}
where
\begin{eqnarray}  
g(\mathbf{p}) = \cos\left({p_x a \over \hbar}\right) - \cos\left({p_y
 a \over \hbar} \right)
\label{gp-d-wave}
\end{eqnarray}
is the $d$-wave form factor. Both theoretical and numerical lattice
models indicate that the richness of the phase diagram results from a
very delicate energy balance between the competing phases and is 
highly sensitive to the parameters of the theory and to external 
variables (e.g., temperature and doping)~\cite{Scalapino06,Yao07}.

Despite the differences in the nature and the scale of their
interactions, QCD and high-$T_c$ materials resemble each other in
significant ways.  We already stressed that chiral symmetry is
restored abruptly as the quark chemical potential is increased just as
antiferromagnetic order vanishes in the presence of a sufficiently
strong magnetic field.  In each case, the varying external field is
not coupled to the order parameter directly but rather 'exhausts'
enough degrees of freedom to destroy the initial order.  For each of
these systems, we allow for the additional possibility of a
thermodynamic competition between pairing and chiral/antiferromagnetic
order.  A major distinction between these two systems is that only two
colors participate in the pairing state of QCD while all spins
contribute to the $d-$wave order parameter of high-$T_c$ materials.

Random matrix models can be constructed as indicated in
section~\ref{s:two-models}, starting with the identification of the
relevant symmetries and a construction of the interactions.  For QCD,
the additional color and spin symmetries need to be included in the
matrix elements: The block matrix $W$ in equation~(\ref{Dirac})
acquires an expanded color and spin block structure and becomes
\begin{eqnarray}
 W = \sum_{\mu=0}^3 \sum_{a=1}^8 \lambda_a\otimes\sigma_\mu\otimes A_{\mu a},
 \label{blockform}
\end{eqnarray}
where $\sigma_\mu=(1,i\mbox{{\boldmath$\sigma$}})$ with
\mbox{\boldmath$\sigma$} the spin Pauli matrices, $\lambda_a$ are the
generators of the color $SU(3)$ algebra, while $A_{\mu a}$ are real
matrices representing single-gluon exchange. As before, ensembles
averages are carried over the gluon field configurations; the matrix
elements of $A_{\mu a}$ are drawn on a Gaussian distribution with zero
mean. To respect Lorentz and color invariances in vacuum, the inverse
variance is chosen to be independent of both indices $\mu$ and
$a$~\cite{VanJac00a, VanJac00b}.

The model for high-$T_c$ superconductivity is constructed in a manner
similar to the phonon-exchange model of section~\ref{ss:ph}.  First, a
parametrization is introduced that partitions the first Brillouin zone
into states related by momentum reversal. In order to accommodate
correlations induced by the AF state of equation
(\ref{AF-OP-microscopic}), a further division is made between states
that are related to one another by a momentum shift of
$\mathbf{Q}=(\pi/a,\pi/a)$. Overall, this gives a total of four
momentum regions, $(\mathbf{p},\omega_n)$,
$(\mathbf{p}+\mathbf{Q},\omega_n)$, $(-\mathbf{p},-\omega_n)$, and
$(-\mathbf{p}-\mathbf{Q},-\omega_n)$, which partition the first
Brillouin zone in four equal parts~\cite{VanJac09}. With this
parametrization, the AF order parameter is written as
\begin{eqnarray}
 \mathbf{m}_{AF}=\left\langle \sum_{r,s=1}^4\sum_{i,j=1}^{M}
 \sum_{\alpha,\beta=\uparrow,\downarrow} \psi^{\dagger}_{r,i,\alpha}
 \mbox{\boldmath$\sigma$}_{\alpha\beta}
 \,\left(\Gamma_{AF}\right)_{r,s}\,\delta_{ij}\,
 \psi^{\phantom{\dagger}}_{s,j,\beta} \right\rangle,
\label{mAF-RM}
\end{eqnarray}
where $r$ and $s$ are region indices and $\Gamma_{AF}$ is the matrix 
\begin{eqnarray}
\Gamma_{AF}= 
\left(
\begin{array}{cccc}
0 & 1 & 0 & 0\\
1 & 0 & 0 & 0\\
0 & 0 & 0 & 1\\
0 & 0 & 1 & 0\\
\end{array}
\right).
\label{GammaAF}
\end{eqnarray}
In constructing the superconducting  order parameter, we 
choose to replace the exact form factor $g$ by the low-energy approximation
\begin{eqnarray}
 \phi_d(\mathbf{p}) = \mathrm{sign}\left(g(\mathbf{p})\right),
\label{simplified-form-factor-d}
\end{eqnarray}
which neglects the detailed momentum dependence of $g$ but 
retains its $d-$wave character by changing sign for every 90 degree rotation of
$\mathbf{p}$.  The partition of the first Brillouin zone can be chosen
so that $g$ is constant in each region~\cite{VanJac09}. The SC order
parameter can then be written as
\begin{eqnarray}
 m_{SC-d} = \left\langle \sum_{r,s=1}^4\sum_{i,j=1}^{M}
 \psi_{r,i,\uparrow} \left(\Gamma_{SC-d}\right)_{r,s}
 \,\delta_{ij}\,\psi_{s,j,\downarrow} \right\rangle
\label{mSC-d-RM}
\end{eqnarray}
where the $d-$wave form factor is given as
\begin{eqnarray}
 \Gamma_{SC-d} = 
\left(
\begin{array}{cccc}
0 & 0 & -1 & 0\\
0 & 0 & 0 & 1\\
-1 & 0 & 0 & 0\\
0 &  1 & 0 & 0\\
\end{array}
\right).
\label{GammaSC-d}
\end{eqnarray}

The construction of the random matrix interactions is not as
straightforward as in QCD, because there is no microscopic theory
which formulates the problem in a form similar to that of the
Yang-Mills action of equation (\ref{QCDaction}).  The problem could
then be approached by writing an effective two-body potential at the
level of the quartic term $Y$ of equation~(\ref{Y-XsB}).  The question
is to identify the most effective two-body potential which accurately
describes the interactions between the lowest-lying degrees of freedom
in the system. This is for instance the approach followed by the
Hubbard or the t-J model. We adopted a quite different approach by
starting at the deeper microscopic level of the action, where we
postulated the existence of a generic interaction carried by the
exchange of either density or spin fluctuations.  In some sense, this
would describe a 'glue' interaction where Lorentz invariance is broken
but rotational symmetry is preserved. Following the spirit of the
models discussed above, no further specifications are made on the
dynamics of interactions, whose matrix elements are drawn again on
Gaussian distributions. A last, but not essential, element of the
construction is the addition in the action of a spin-independent term
$\psi^\dagger \Gamma_t \psi$ with
\begin{eqnarray}
 \Gamma_t =  \textrm{diag}(t,-t)\otimes\left(
\begin{array}{cccc}
1 & 0 & 0 & 0\\
0 & -1 & 0 & 0\\
0 & 0 & 1 & 0\\
0 &  0 & 0 & -1\\
\end{array}
\right),
\label{Gammat}
\end{eqnarray}
which mimics nearest neighbor hopping. Diagonal terms can be positive
or negative, as are the free electron energies
$\varepsilon(\mathbf{p}) = - 2 t_0 (\cos(p_x a /\hbar)+\cos(p_y a
/\hbar))$. Similar to the nesting condition
$\varepsilon(\mathbf{p}+\mathbf{Q}) = - \varepsilon(\mathbf{p})$, the
elements of $\Gamma_t$ change sign under a change of region corresponding
to a momentum shift $\mathbf{Q}$~\cite{VanJac09}.

As constructed, the QCD and high-$T_c$ models can be solved exactly in
the thermodynamic limit by following the methods described in
section~\ref{s:two-models}. Not surprisingly, the associated
thermodynamic potentials have a very similar structure. Working again
with the smallest Matsubara terms, we find the generic form
\begin{eqnarray}
 \Omega(\sigma,\Delta) = A |\Delta|^2 + B \sigma^2 - \sum_{\ell,\pm}
 \log(E_{\ell\pm}^2+\pi^2T^2),
\label{pot-both}
\end{eqnarray}
This result is quite remarkable: Even though underlying microscopic
physics is very different in each system, the resulting potentials
share a common structure. In fact, the models simplify greatly as the
calculations proceed.  Starting from a complicated block structure and
many different variances at the level of the action, the final result
depends only on the two constants $A$ and $B$, which scale linearly
with the variances~\cite{VanJac00a,VanJac00b, VanJac09}. In equation
(\ref{pot-both}), $\Delta$ are pairing fields while $\sigma$ denotes
either the chiral or the AF field.  The $E_{\ell\pm}$ are the
quasifermion energies in a fixed background of $\sigma$ and
$\Delta$. In QCD, the sum over $\ell$ is carried over the three
colors. As only two colors participate in pairing, we have two
\emph{ungapped} excitations of energy $E_{\pm} = \pm (\sigma + m) -
\mu$ (as before in equation~(\ref{OmegaRMCsB})) and four \emph{gapped}
excitations of energy $E_\pm = [((\sigma +
  m)\mp\mu)^2+|\Delta|^2]^{1/2}$. For high-$T_c$, all excitations are
gapped and have an energy $E_\pm =
[(\sqrt{t^2+\sigma^2}\mp\mu)^2+|\Delta|^2]^{1/2}$. These expressions
are similar to the quasiparticle energies that were found in the
microscopic theory of Ref.~\cite{Kyung2000}, and which can be written
as $E_\pm(p)=[(\sqrt{\varepsilon(p)^2+\sigma^2}\mp\mu)^2+g(p)^2
  |\Delta|^2]^{1/2}$, where $g(p)$ is the form factor given in
(\ref{gp-d-wave}).  Approximating $\varepsilon(p)$ by $\pm t$ and
$g(p)$ by $\phi_d(p)$ as in (\ref{simplified-form-factor-d}), so that
$g^2\approx \phi^2 = 1$, the quasiparticle energies of the microscopic
theory reduce to those of the random matrix theory.

The phase structures are readily derived from the two gap equations
$\partial\Omega/\partial \sigma = 0$ and $\partial\Omega/\partial
\Delta = 0$, which are polynomial in $\sigma$ and $\Delta$ and can
thus be solved exactly.  It is also interesting to note that it is
possible to rescale all fields and external parameters by a
multiplicative constant, with the result that the phase topology
depends only on the single ratio of $\alpha = B/A$. In QCD
(high-$T_c$), small values of $\alpha$ favor chiral symmetry breaking
(AF order) and large ratios favor superconductivity.

For QCD, an interaction mediated by single-gluon
exchange has a ratio $B/A=3/4$. The corresponding phase diagram is
shown in figure~\ref{f:QCD-pd}-(b). In addition to the chiral phase
structure obtained in the model of section~\ref{s:two-models}, a
superconducting phase (SC) is found at the low-temperature/high
density end of the plot. The transition is first order as the chiral
broken (CB) and SC phases cannot be directly related to one another by
a continuous transformation.  It also is interesting to see how the
phase structure evolves when $\alpha$ is varied away from the value
$\alpha=3/4$ associated with QCD.  In total, only six different
topologies can be realized as shown in figure~\ref{f:QCD-pd} and the
phase structure of figure~\ref{f:QCD-pd}-(b) is found over a large
range of values of $\alpha$, i.e. $0.42 < \alpha < 1.05$. It can thus
be concluded that this phase structure, which we associated with QCD,
is fairly robust against moderate variations in the detailed
description of the random matrix interactions. The random matrix
results are consistent with those of early microscopic models of color
superconductivity~\cite{Alford1998,Berges1998}. Some microscopic
models also found a wedge of mixed broken symmetry extending over the
small temperature end of the first-order line, as shown in
figure~\ref{f:QCD-pd} (c)~\cite{Kizatawa02,Hatsuda06}. Note, however,
that this phase structure is observed in the present model over a
fairly small range of values of $\alpha$, i.e. $1.05<\alpha<1.06$. It
thus appears to be more fragile than the structure of
figure~\ref{f:QCD-pd} (b).  The robustness of the coexistence phase is
also discussed in the context of NJL models in~\cite{Buballa05}.

 \begin{figure}
   \begin{center}
     \begin{tabular}{cc}
      \includegraphics[width=0.48\textwidth]{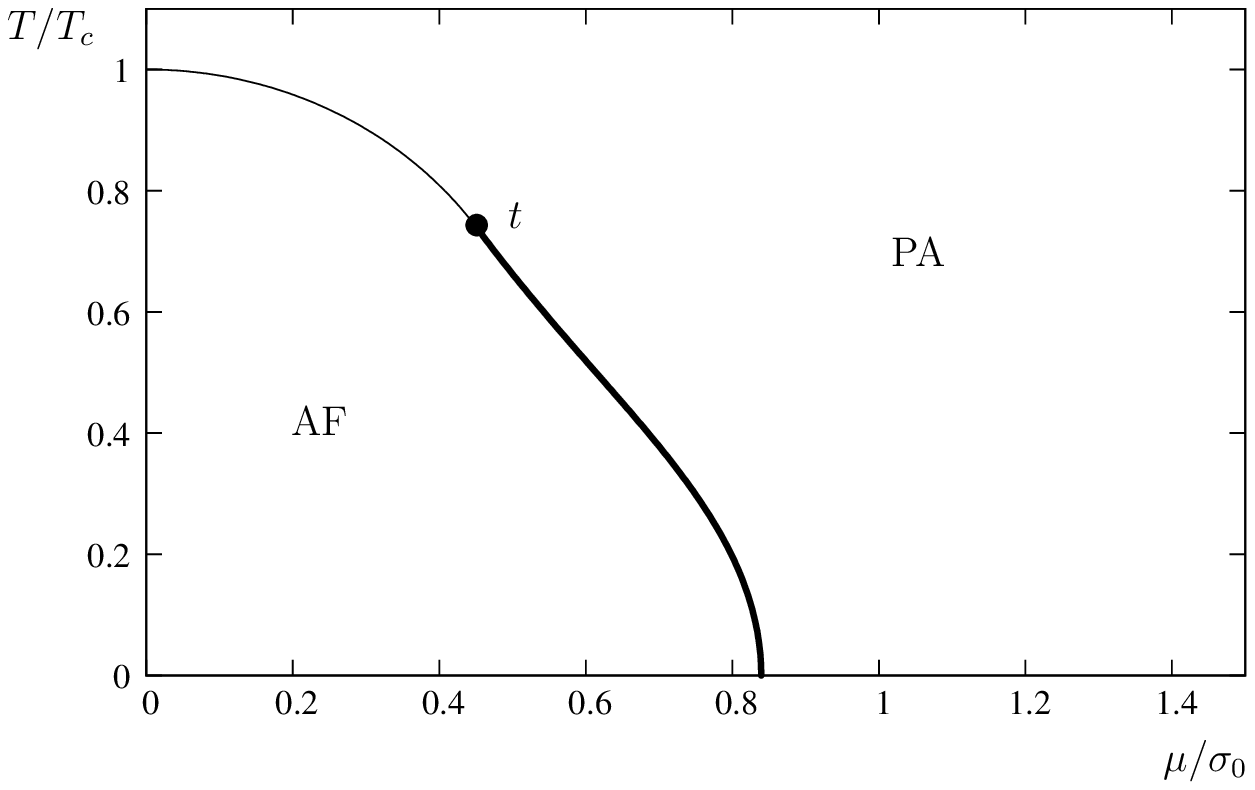} &
     \includegraphics[width=0.48\textwidth]{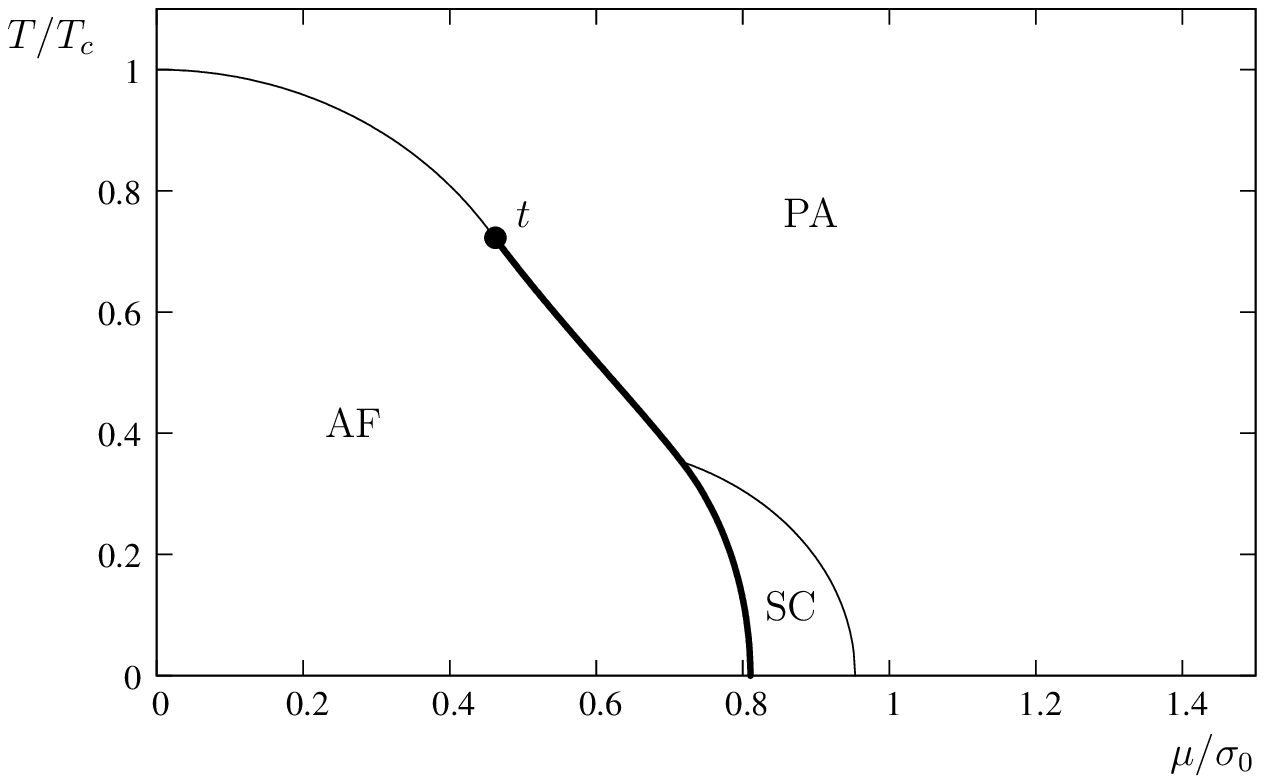} \\
      (a) & (b) \\
      \\
     \includegraphics[width=0.48\textwidth]{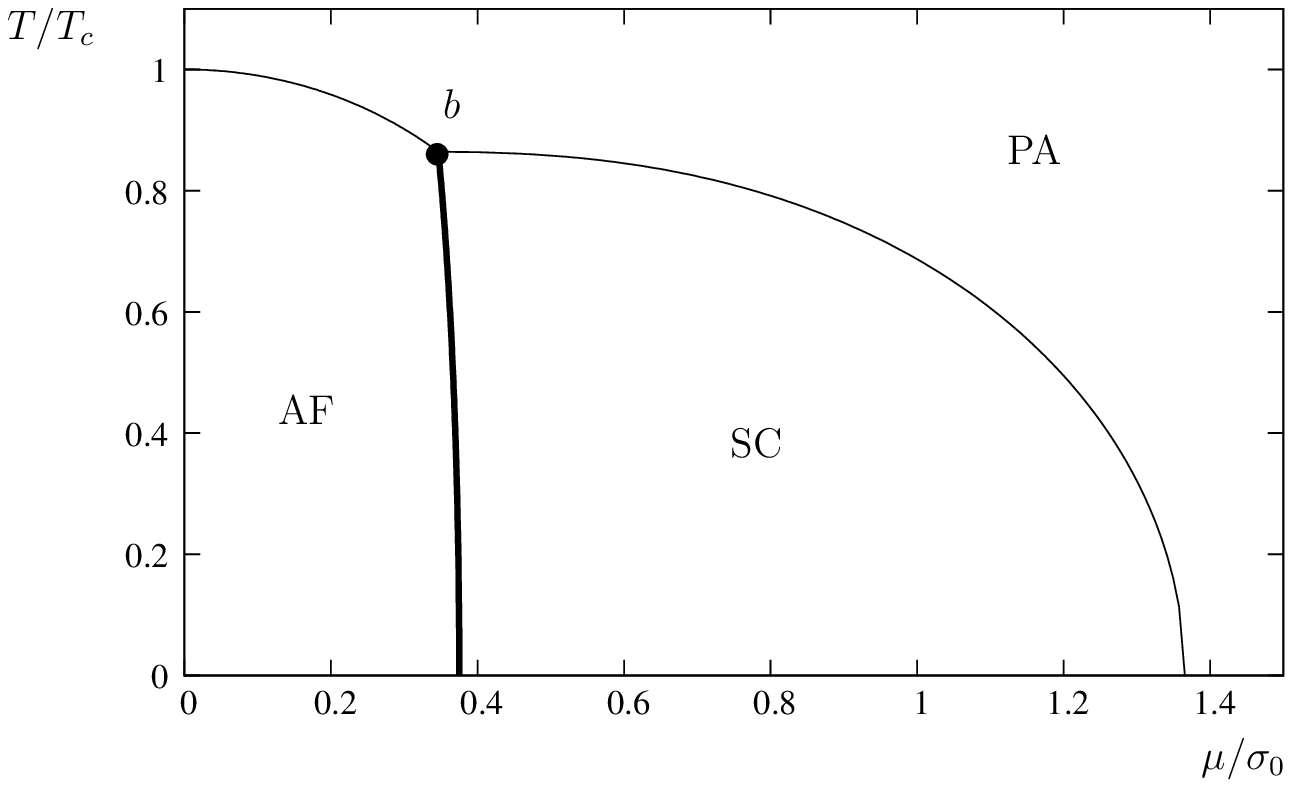} &
     \includegraphics[width=0.48\textwidth]{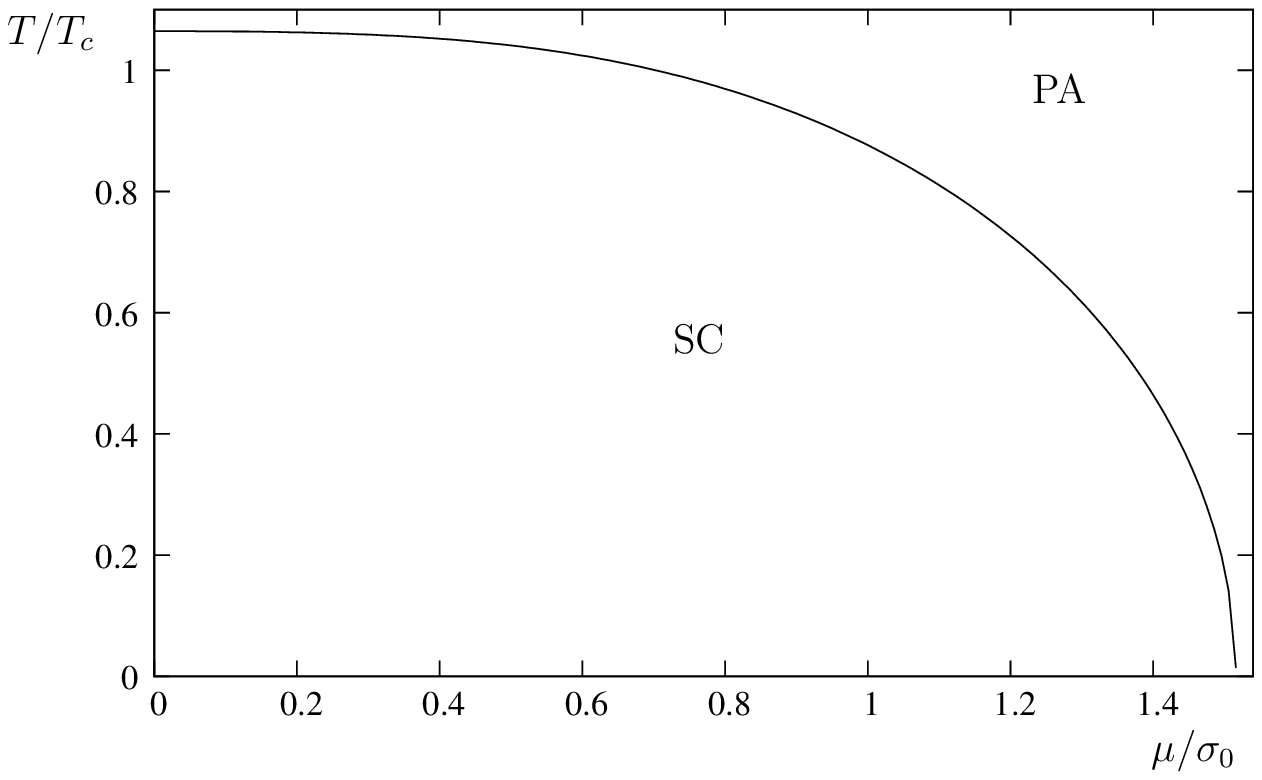} \\
     (c) & (d) \\
     \\
     \end{tabular}
\caption{The four allowed topologies for the phase diagram of
  high-$T_c$ models. AF is the antiferromagnetic phase, PA the
  paramagnetic one, and SC the superconducting phase.  First- and
  second-order transitions are depicted by thick and thin lines,
  respectively. $t$ is a tricritical point and $b$ is a bicritical
  one. $T_c$ is the transition temperature for $\mu=0$ and small
  $\alpha$ and $\sigma_0$ is the AF field at zero $\mu$. A hopping
  term $t = 0.5 \sqrt{2/B}$ is assumed. From top left to bottom right,
  the panels correspond to $\alpha=0.07$, $\alpha=0.2$, $\alpha=0.8$,
  and $\alpha=1.1$. The figures are adapted
  from~\cite{VanJac09}.\label{f:highTc-pd} }
   \end{center}
 \end{figure}

A similar investigation can be performed for the high-$T_c$ model.
Here, in the absence of a detailed theory at the level of the action,
the value of $\alpha$ is not known a priori. In order to restrict the
size of the parameter space to be explored, we make a number of
assumptions~\cite{VanJac09}: First, an AF state is expected at
half-filling $\mu=0$. This leads us to force fluctuation exchanges to
be stronger for spin than for density.  Second, we consider
interactions that are attractive in the $d-$wave channel but repulsive
for $s-$wave pairing. This requires interactions that
 favor large momentum exchange of order $\sim\mathbf{Q}$. The two
 desired properties can be realised simultaneously with an
 appropriate choice for the variances of the random matrix
 interactions~\cite{VanJac09}. For a given material, the variances
 must reflect the different strengths of the scattering elements
 between the four regions of states in the first Brillouin zone. Once
 the two conditions above have been imposed, the variances can still
be varied over a continuous --- but restricted --- range of
 values.  The resulting phase structure depends on the single
 parameter ratio $\alpha$, which can assume any positive value. The
 many allowed possibilities for the variances reduce in this case to
 only four distinct phase topologies, which are illustrated in
 figure~\ref{f:highTc-pd}.

How well do these phase structures compare to experiments?  High-$T_c$
cuprates exhibit an insulating AF phase at half-filling ($\mu=0$),
with an AF ordering that weakens as a function of doping and
eventually disappears in favor of an SC phase with a domelike phase
boundary (See, e.g.,~\cite{Lee2006, Damascelli2003,
  Demler2004}). Figures~\ref{f:highTc-pd} (b) and (c) already exhibit
the coarse structure of this picture. However, a closer analysis
reveals several discrepancies. First, the present random matrix
approach does not distinguish electron- from hole-doped
materials. This can be traced back to the kinetic energy term, which
takes only nearest-neighbor hopping into account and exhibits the
nesting property. As a result, the thermodynamic potential is an even
function of $\mu$. In principle, next-nearest neighbor and
higher-order hopping terms could be added to the model by using a
hopping matrix whose eigenvalues have a sign distribution that
reproduces the symmetries of the corresponding kinetic energies in the
various regions of the Brillouin zone. Although we have limited
ourselves to nearest-neighbor terms here, we expect that a model with
a more elaborate hopping matrix would distinguish between electron and
hole dopings. The second discrepancy concerns the transition between
AF and SC phases, which is reported to occur either through phase
separation, structured spin and charge distributions, or via a mixed
AF+SC phase.  (E.g., see the discussion in~\cite{Demler2004} and
references therein.) Here, only the first scenario of a first-order
transition is found in all cases. A third discrepancy can be observed
in the absence (for hole doped materials) of a pseudogap region
separating the AF and SC phases. We will make further comments on the
last two discrepancies in the end of this section. 

Returning to the similarities between the random matrix models for QCD
and high-$T_c$ materials, limitations regarding the extension of some
phases can be found in both cases. In QCD, a microscopic theory would
contain the same logarithmic singularity as that present in the right
side of equation (\ref{gapeqph}) so that the diquark phase would
extend to arbitrarily high $\mu$. In the high-$T_c$ model, a similar
singularity exists in the gap equation for AF~\cite{Hirsch85,Lin87}
and results in an antiferromagnetic phase that exists independent of
the hopping strength $t$. In the random matrix model, by contrast, $t$
must be kept below a threshold value for the AF to be
realized~\cite{VanJac09}.  As discussed earlier in
subsection~\ref{ss:ph}, these discrepancies are acceptable in a
mean-field approach as they concern situations where the condensates
are weak and thus fragile.

A striking difference can be observed in the thermodynamic competition
between the order parameters when comparing the diagrams in
figures~\ref{f:QCD-pd} and~\ref{f:highTc-pd}. Both types of systems
support the coexistence of different phases along a first-order line.
The two competing phases then coexist at a given temperature with
different densities. They form an inhomogeneous state whose detailed
spatial structure depends on the energetic cost of creating an
interface.  Clearly, such a description is beyond the simple approach
considered here.  A completely different situation occurs for the
quark systems only.  As $\alpha$ is increased in favor of
color-superconductivity (with values that are no longer representative
of QCD), a tetracritical point --- the meeting point of four different
continuous transition lines --- appears. The transition from one phase
to the other then proceeds continuously through an intermediate phase
of mixed broken symmetry, where both orders coexist in a spatially
homogeneous phase. For the high-$T_c$ system, such a mixed broken
symmetry state is absent. At best, a bicritical point appears, marking
the end of a first-order line with its discontinuous transition.

The absence of a tetracritical point in the high-$T_c$ case is a
direct result of the absence of ungapped excitations.  The conditions
for the observation of a tetracritical point can be determined by expanding
the thermodynamic potential of equation (\ref{pot-both}) as a power series
\begin{eqnarray}
\Omega(\sigma,\Delta) \approx \Omega(0,0) + a_2 \sigma^2 + b_2 |\Delta|^2 + 
a_4 \sigma^4 + b_4 |\Delta|^4 + c_4 \sigma^2 |\Delta|^2 + \ldots, 
\end{eqnarray}
where the coefficients are known functions of $\mu$ and $T$. The AF
state makes a transition to a paramagnetic state along the line $a_2 =
0$, whereas the superconducting gap vanishes along the second-order
line $b_2 = 0$. A critical point arises at the crossing of those two
lines.  A mixed broken symmetry phase (with non-zero values of both
$\sigma$ and $\Delta$) can be stable near this critical point provided
that the coefficients of the quartic terms satisfy the inequality
$\rho_4 = 4 a_4 b_4 - c_4^2 > 0$~\cite{VanJac09}.  In this case, the
critical point is a tetracrictical one.  In the opposite case, $\rho_4 <
0$, the critical point is only bicritical.  For the high-$T_c$ model,
it can be shown that $\rho_4 < 0$ for all values of $\alpha$. If a
small fraction, $f$, of the excitations were forced to be
ungapped, $\rho_4$ would become positive for a range of values of
$\alpha$ near unity, and a tetracritical point would be present for
those couplings.

Although the
mixed broken symmetry phase is never realized as a global minimum of
the thermodynamic potential of the high-$T_c$ model, it is always an
unstable solution of the gap equations, i.e., a saddle-point of
$\Omega(\sigma,\Delta)$. Further, the difference in energy between this
state and the global minimum of $\Omega(\sigma,\Delta)$ is small on the
scale of the inverse variance. This suggests that the existence of a
mixed broken symmetry phase is a delicate matter.  It may well occur
in specific dynamic models or in calculations more detailed than those
presented here, e.g., calculations beyond mean-field level, but the
demonstration that its appearance is robust may be challenging.  This
is an example of useful information that can be obtained from a random
matrix approach even though a precise assessment of those effects
requires more than the mean-field treatment of random matrix theory.
 
 A question related to the occurrence of critical points is whether
 the theory exhibits, approximately or exactly, condensation patterns
 of higher symmetry near these points.  This would be the case, for
 instance, if the series expansion of $\Omega(\sigma,\Delta)$ could be
 expressed as a polynomial in a single variable, i.e., a linear
 combination of $\sigma^2$ and $\Delta^2$.  In such a case, the
 different phase transitions could be described as a rotation of a
 metavector order parameter constructed with both $\sigma$ and
 $\Delta$, in the manner of the SO(5) theory of superconductivity of
 Zhang~\cite{Zhang1997} or the O(10) theories that have been proposed
 for QCD~\cite{Wiese2000}.  We have found no indications of such higher
 symmetry in the vicinity of a critical point for either the quark or
 the high-Tc systems described here.  There is one interesting case:
 In the high-$T_c$ model in the limit $\mu =0$, the potential of
 equation (\ref{pot-both}) becomes a function of the combination
 $\sigma^2 + \Delta^2$ provided that $\alpha = B/A = 1$.  This higher
 symmetry is manifest only at $\mu=0$ and is broken for $\mu >0$ in
 favor of an SC ground state.

We end this discussion with some comments on possible extensions of
the model for high-$T_c$ superconductors. The present construction is
based on only a few elements, and we have shown how it can help in
identifying symmetry-related results. At the same time, the model
cannot be expected to describe the entire spectrum of materials, which
can greatly differ in their detailed microscopic interactions. We have
noted that some of this variety can be encoded in the choices for the
variances to be associated with scattering processes between the
different states of the first Brillouin zone. The strength of these
matrix elements can most likely be probed by two-point correlation
functions such as spin or density susceptibilities, which are
experimentally accessible. Further work is needed in this
direction. We also mentioned earlier the lack of a pseudogap phase in
the random matrix phase diagram.  A pseudogap region is thought to
arise as a result either of correlations competing with the AF and SC
orders or from preformed Cooper pairs leading to large phase
fluctuations in the pairing field~\cite{Lee2006}.  We have not
attempted to model the pseudogap phenomenon here but can offer
tentative indications of the effort that would be required in this
direction: The scenario of a competing order could in principle be
taken into account by decomposing the four-fermion random matrix
potential $Y$ simultaneaously into the AF, SC, and the pseudogap
channels (provided one can make meaningful ``sign approximations''
similar to those of the SC channel) and by constructing a new
thermodynamic potential. The difficulties that could arise in this
approach would be the lack of a true density of states distributed
around a Fermi sea. This would make it difficult to obtain a faithful
description of the Fermi arc phenomenology but might still capture
some of the interesting physics. The scenario of phase fluctuations is
less easily addressed in the present approach, since the degrees of
freedom associated with the fluctuations need to be introduced
explicitly in the model.

\section{Extension to a model for the ferropnictides}
\label{s:oxy}

As an example of a possible extension of the models discussed so far,
we turn to the recently discovered ferropnictides, which constitute
another class of type-II superconducting materials exhibiting magnetic
and superconducting correlations. A current discussion of the iron
arsenide compounds concerns the symmetry of the pairing state and its
relationship to the phase diagram.  It has been claimed by several
authors~\cite{Fernandes2010,Fernandes2010a,Vavilov2010,Vorontsov2010}
that the occurrence of a mixed broken symmetry phase would constitute
an evidence for a pairing with an $s_{+-}$ symmetry, as opposed to the
conventional $s_{++}$-wave.  This claim relies on the observation that
$s_{+-}$ pairing leads to a mixed broken symmetry state if the two
following conditions are met: The system has a finite chemical
potential breaking the particle-hole symmetry, and the electron band
is elliptic whereas the hole band is circular.  In contrast, an
$s_{++}$ state meeting these two conditions does not lead to a mixed
broken symmetry state~\cite{Fernandes2010,Fernandes2010a}.

The main elements of the corresponding microscopic theory are the
following~\cite{Mazin2009}: The electronic structure of the system
shows two hole pockets at the center of the first Brillouin zone and
electron pockets shifted at either $\pm (\pi,0)$ or $\pm (0,\pi)$ (for
a square lattice with one Fe atom per cell).  An AF state may develop
with a form similar to equation (\ref{AF-OP-microscopic}), where the
sum over states $\{\mathbf p\}$ is now carried over one type of
pocket, whereas the corresponding $\{{\mathbf p} + {\mathbf Q}\}$ states with
${\bf Q} = (\pi ,0)$ belong to the other type of pocket.  In the
$s_{+-}$ pairing state, the gap function changes sign from the hole
pockets to the electron pockets, whereas an $s_{++}$ gap function
assumes the same sign in all pockets.

It is of interest to see how the model presented in the previous
section should be modified to describe such
materials. Following~\cite{Fernandes2010a}, we assume two electronic
bands shifted by the momentum ${\bf Q}$ with respect to each other.
One can easily partition the first Brillouin in four equivalent
regions related to one another by a shift in $\mathbf{Q}$ or by
momentum reflection.  The AF correlator then takes the same form
given in equation (\ref{GammaAF}).  Similarly, the sign change that is
characteristic of an $s_{\pm}$-wave state turns out to be represented
correctly by the matrix in (\ref{GammaSC-d}).  Finally, by collapsing
each of the electron and hole bands into a single energy term, e.g.,
$t_1$ for electrons and $-t_1$ for holes, the corresponding
single-fermion Hamiltonian is now described by a term
$\psi^{\dagger}\Gamma_t\psi$, where $\Gamma_t$ is given in
(\ref{Gammat}) with the term $\textrm{diag}(t,-t)$ replaced by a
simple multiplicative factor, $t_1$.

By postulating random matrix interactions of the same form as in the
model of the previous section, the resulting thermodynamic potential
is precisely that of equation (\ref{pot-both}), for which all
excitations are gapped. Thus, the phase diagrams are the same as those
for the cuprates, and the random matrix model would preclude the
occurrence of a mixed broken symmetry state. This is not inconsistent
with the conclusions of \cite{Fernandes2010,Fernandes2010a} since only
one of two required conditions are met by the random matrix
interactions. Here the model ignores the ellipticity of the electron
bands. However, according to the discussion of the previous section, a
mixed broken symmetry state might arise if some of the excitations do
not participate in pairing, i.e. a fraction of the excitations are
ungapped.

Two more results can be derived: First,
following~\cite{Fernandes2010a} and assuming equal coefficients for
the quadratic terms of equation (\ref{pot-both}), i.e., $A = B$, one
finds that the critical point $b$ is located on the $\mu=0$ axis and has
a {\it vanishing} coefficent $\rho_4=4 a_4 b_4 - c_4^2$. This point is
thus at the verge of becoming a tetracritical one. In contrast,
assuming an $s_{++}$ state and fixing $A=B$ yields a critical point
which has $\rho_4 < 0$ and is thus bicritical. These results are consistent
with the microsocopic model of~\cite{Fernandes2010a}.

The above findings demonstrate that the random matrix model already
captures many elements of the physics contained in the microscopic
model of~\cite{Fernandes2010a}. It will be interesting to see whether
the ellipticity of the electron states might also be mimicked in the
random matrix approach and how it might affect the results.

\section{Conclusions}
\label{s:conclusions}

\subsection{A summary}
\label{ss:synthesis}

We have reviewed the construction of random matrix models for the
phase diagrams of a variety of systems.  The general strategy starts
at the level of the action with a description of the interactions
between fermions and exchange fluctuation fields.  These interactions
assume the form of matrices with a global block structure dictated by
the symmetries of the theory.  The detailed form of the interactions
is simplified materially by drawing matrix elements in the individual
blocks at random.  The result is a mean-field model which is exactly
soluble. The corresponding thermodynamic potential has a simple
structure involving only a minimal number of parameters and an
explicit sum over Matsubara frequencies.  This form is perfectly
suitable for numerical purposes, but even greater simplification can
be obtained by truncating this sum to a single term.  While this
truncation has quantitative consequences, it does not alter the
topology of the resulting phases for the cases considered.

Despite --- or even because of --- the many simplifying approximations
made, we believe that this approach to phase diagrams can be of value.
In many calculations, it can be difficult to distinguish between solid
results that are protected by the underlying symmetries of the problem
and more tentative result that depend on the specific model
investigated.  By performing an ensemble average over theories, the
results of the random matrix approach are dictated by symmetries
alone.  They can thus provide guidance in identifying model-dependent
and model-independent features.  Similarly, they can be useful in
understanding the minimal number of symmetry constraints required to
reproduce specific phase structures.  The great simplicity of the
thermodynamic potential obtained in these calculations makes it easy
to check the robustness of predictions with respect to variations in
the description of the interaction.  We also note that near critical
points, random matrix models bear strong similarities to
Ginsburg-Landau theories with the advantage of the presence of
additional constraints, which are inherited from the symmetries of the
random matrix interactions.  The freely-adjustable parameters of
Ginsburg-Landau theory are replaced by known functions of temperature
and chemical potential.  As we have demonstrated above, these
constraints can be extremely helpful in ruling out certain topologies
in the phase diagram.

Difficulties associated with the thermodynamic limit can be formidable
when calculating the phase diagram.  The analysis in the previous
sections showed that the partition function is exponentially small in
the volume of the system.  Moreover, the thermodynamic limit does not
commute with other limits in the system. As a result, care must be
taken in cases where low-energy metastable states exist, as their
contribution to the partition function may be difficult to be
distinguished from that of the ground state.  In such circumstances, a
random matrix model can be useful in identifying the metastable
states. For systems with a finite fermion chemical potential, the
situation can be significantly worse.  For example, the QCD Dirac
operator is complex as a consequence of the fermion sign problem, and
numerical calculations with traditional sampling methods are
precluded.  In such cases, random matrix methods represent one of the
few options available.

In the form presented here, random matrix models can be extended to
any system where the fermions are the main degrees of freedom, the
symmetries of the interaction are central to the problem, and the
detailed dynamics of the exchange fields plays a secondary role. The
model can be realized in either of two ways: The simplest approach
consists of constructing the thermodynamic potential directly. This
requires the identification of the auxiliary fields to be associated
with the order parameters of the system and the determination of the
single fermion energies in a fixed background of those fields. The
remaining parameters could then be chosen freely.  The richer approach
adopted here is to start at the deeper level of the action.  The
reward for this extra effort is insight regarding the constraints that
are imposed as a consequence of the symmetries of the interactions and
their effect on the global properties of the system.

\subsection{Open questions}
\label{ss:open}

The various examples presented here indicate that a random matrices 
offer an interesting approach to the construction of phase diagrams.  
However, a number of open question remain:
\begin{itemize}
\item 
  {\bf Fermion sign problem}.  Although the random matrix model can be
  solved at finite chemical potential --- using a saddle-point method
  which becomes exact in the thermodynamic limit --- numerical
  calculation of the same model are plagued by the fermion sign
  problem.  Difficulties present at the microscopic level apparently
  disappear when working at a macroscopic level.  The open question is
  thus to understand the resolution of this apparent contradiction and
  to see whether this mechanism can be exploited to ease the fermion
  sign problem.  Considerable work in this direction has been
  performed in a QCD context (see, e.g.,
  \cite{fs0,fs1,fs2,fs3,fs4,fs5}), where an interesting
  deterministic pattern in the sign oscillations of the fermion
  determinant has been seen.
\item
  {\bf The nature of the interactions at the microscopic
  level}. The random matrix model for high-$T_c$ was inspired by
  earlier models for QCD.  Given this background, it was natural to
  identify the interactions at the more microscopic level as being
  mediated by the exchange of density and spin fluctuations. Is this
  structure generic, and can it represent other kinds of interactions,
  e.g., a static superexchange?
\item
  {\bf Eigenvalue correlation for SC and AF states}.  In
  section~\ref{ss:chiral} we noted that correlations due to chiral
  symmetry in QCD could be analyzed either at the microscopic level of
  the Dirac spectrum or at the level of the phase diagram. A similar
  analysis has not been performed either for the pairing state in QCD
  or for the AF and SC states in high-$T_c$ superconductivity.  It
  would be of interest to see if the symmetries underlying these
  states induce similar microscopic correlations for a suitable single
  particle operator.  Such an analysis might require the construction
  of an analogue of the Banks-Casher relationship.
 \item
  {\bf Summations over Matsubara frequencies}.  Since our primary
  concern has been to determine the physically realizable topologies
  of phase diagrams, we were willing to make simplifications in the
  thermodynamic potential which preserved their topology.  This led us
  to truncate the infinite sum over Matsubara frequencies given by
  (\ref{OmegaRMCsB}) by a single term.  The resulting form, given by
  (\ref{OmegaSimple}), is sufficiently simple that it can be studied
  `by hand', but it is quantitatively different from the full result.
  It would be interesting to determine the extent to which the full
  random matrix result of (\ref{OmegaRMCsB}) can provide a
  quantitative description of model systems for which the phase
  diagram can be constructed reliably. A related question is whether
  the inequality $\rho_4 \leq 0$ for all $\alpha$, found for both
  high-$T_c$ and iron arsenide materials, holds for the full result. A
  direct comparison with the microscopic theory
  of~\cite{Fernandes2010a} shows that $\rho_4$ has the same sign in
  both the random matrix approach and in the microscopic theory when
  $\alpha$ is near $1$.  It would be interesting to see if this result
  remains valid in other cases.

\item
  {\bf Why do random matrix models work?}  We end this discussion
  by offering some admittedly speculative suggestions about the origin
  of the success of random matrix theory.  Consider a real gas that
  obeys the ideal gas law.  Interactions between its molecules are
  strong, but their role is strictly limited to the establishment of a
  statistical population of the states.  The properties of the gas
  then follow from purely statistical arguments that are independent
  of the nature of the underlying molecular interactions.  (Clearly,
  no one would consider attempting to determine the details of the
  molecular interaction from the ideal gas law.)  Similar ideas form
  the basis of Niels Bohrs' compound nucleus model~\cite{NielsBohr}.
  It is presumed that the energy of an incoming particle is rapidly
  distributed statistically among the constituent nucleons.  The
  description of the eventual decay of the excited state is determined
  solely by this statistical distribution and is independent of the
  details of the formation process.  This is directly related to
  Wigner's ideas that random matrix theory should apply whenever
  single-particle energy levels are strongly intermixed and contribute
  in an equilibrated or `democratic' way to physical observables.  To
  the extent that these notions can be extended to encompass the
  thermodynamic properties of QCD or high-$T_c$ materials, the
  decision to replace the exact action by an ensemble average should
  be legitimate.  The phase diagram inherits only the symmetries of
  the underlying action and is independent of its detailed properties.

Two additional clues point to the importance of strong energy level
mixing.  In 1984, Bohigas, Giannoni, and Schmit observed that the
level correlations of the Sinai billiard --- a quantum system with a
chaotically behaved classical analog --- were consistent with the
predictions of RMT (in this case, those of the Gaussian orthogonal
ensemble).  They further conjectured that RMT should correctly predict
the spectral correlators of any system whose classical analog is
chaotic~\cite{BohGiaSch84}. Here, QCD and high-$T_c$ superconductivity are
both described by non-linear field theories.  The corresponding
classical theories, and even more so the quantum many-body theories,
are expected to be chaotic (i.e., ergodic) in nature.

A second indication for the need for energy mixing comes from observed
limitations on the range of validity of the RMT description of
spectral correlators.  In principle, RMT is expected to provide an
exact description of all spectral correlators on an energy scale which
is vanishingly small compared to the full support of the spectrum.  In
practice, RMT predictions for many systems (including QCD) agree for
eigenvalues with are smaller than the so-called Thouless energy,
$E_{\mathrm{Th}}$~(e.g., see~\cite{review2}).  This scale can be
determined using ideas from localization theory, where
$E_{\mathrm{Th}}$ is inversely proportional to the time required for a
particle to diffuse across the system.  For energies greater than
$\sim E_{\mathrm{Th}}$, a particle crosses the system in a time that
is too short to resolve individual states. Given that the reflections
at boundaries scatter the particle to other energy states, a large
number of individual energy states end up contributing to the wave
function, thus yielding the apparent 'democracy' underlying the RMT
description.

Ultimately, a strong argument for the value of the random matrix
approach to phase diagrams should rely on clues for universality on
both microscopic and macroscopic scales. For a chiral random matrix
model of QCD at zero $\mu$ and with a temperature dependence given by
the two lowest Matsubara frequencies, it was shown that the
thermodynamic approach and the Banks-Casher relationship yield exactly
the same chiral condensate as a function of
temperature~\cite{Jackson1997}.  These results strongly suggest that
the physical content of the thermodynamic approach is actually
dominated by the microscopic spectral density --- which we believe to
be universal. For $\mu>0$ and $T=0$, it was strongly argued that a
random matrix model with the global symmetries of QCD yield a
microscopic spectral density of the same form as that of the QCD Dirac
operator~\cite{Osborn2004,Akemann2005}. If similar connections can be
established over most the ($\mu$,$T$) plane and extended to a
temperature dependence that includes all Matsubara frequencies, then
one can expect that the phase diagrams that we constructed can be
justified on the basis of the universality of the corresponding
microscopic correlators. It would be of particular interest to see if
there is a relation between the inequality for $\rho_4$ identified for
the phase diagrams of superconducting materials and the properties of
the microscopic correlators. This also indicates that there is
considerable interest in studying the microscopic spectral properties
of condensed matter systems!
\end{itemize}

One final open question remains: Why is the random matrix approach
described here so rarely used in the study of phase diagrams?  The
answer probably lies in the fact that the starting form of the random
matrix can be dauntingly complicated.  For example, the block form of
the random matrix appropriate for the description of the color and
spin symmetries of QCD~(\ref{blockform}) is $6\times 6$ and contains
$32$ component matrices.  As we have attempted to explain, these complications
disappear step-by-step as one follows the relatively direct path
towards the thermodynamic potential of (\ref{OmegaRMCsB}) and the even
simpler (but topologically equivalent) form of (\ref{OmegaSimple}).
Given the model-independence of the final phase diagram and the merits
of this approach relative to Ginsburg-Landau theory as noted above, we
believe that the initial investment in constructing the random matrix
is richly rewarded and can be recommended.

\end{document}